\documentclass[useAMS,usenatbib]{mn2e}
\pdfoutput=1
\usepackage{graphicx}
\usepackage{subfigure}
\usepackage{natbib}
\usepackage[english]{babel}
\usepackage{amssymb}
\usepackage{amsmath}
\usepackage{xspace}
\newcommand\ionm[2]{#1$\,${\small\rmfamily{#2}}} 
\newcommand{\hi}{\ionm{H}{I}\xspace}
\def\hii{\ifmmode {\mbox H{\scshape ii}}\else H{\scshape ii}\fi\xspace}
\def\h2{\ifmmode {\mbox H$_2$}\else H$_2$\fi\xspace}
\def\frach2star{$\frac{\rm{M}_{\rm H2}}{\rm{M}_{\star}}$}

\title[CO and CII luminosity functions out to z $=$ 6]{Sub-mm Emission
  Line Deep Fields: CO
  and [CII]  Luminosity Functions out to z $=$ 6}
\author[G. Popping et al.]{Gerg\"o Popping$^{1}$\thanks{E-mail:
    gpopping@eso.org}, Eelco van Kampen$^{1}$, Roberto Decarli$^{2}$,
  Marco Spaans$^{3}$, \newauthor Rachel S. Somerville$^{4}$, Scott
  C. Trager$^{3}$\\
$^{1}$European Southern Observatory, Karl-Schwarzschild-Strasse 2,
85748, Garching, Germany\\
$^{2}$Max-Planck Institut f\"ur Astronomie, K\"onigstuhl 17, D- 69117, Heidelberg, Germany\\
$^{3}$Kapteyn Astronomical Institute, University of Groningen, Postbus 800, NL-9700 AV Groningen, the Netherlands\\
$^{4}$Department of Physics and Astronomy, Rutgers University, 136 Frelinghuysen Road, Piscataway, NJ 08854, USA\\
}

\begin{document}

\maketitle

\begin{abstract}
Now that ALMA is reaching its full capabilities, observations of sub-mm
emission line deep fields become feasible. We couple a semi-analytic model of galaxy formation
with a radiative transfer code to make predictions for the luminosity
function of CO J$=$1--0 out to CO J$=$6--5 and [CII] at redshifts z$=$0--6. We find that: 1) our model
correctly reproduces the CO and [CII] emission of low- and
high-redshift galaxies and reproduces the available constraints on the
CO luminosity function at $z \leq 2.75$;
2) we find that the CO and [CII]
luminosity functions of galaxies increase from $z=6$ to $z=4$, remain
relatively constant till $z=1$ and rapidly decrease towards $z=0$. The
galaxies that are brightest in CO and [CII] are found at $z\sim2$; 3) the CO J$=$3--2 emission line is most favourable to study the CO
luminosity and global \h2 mass content of galaxies, because of its
brightness and observability with currently available
sub-mm and radio instruments; 4) the luminosity functions of high-J CO
lines show stronger evolution than the luminosity functions of low-J
CO lines; 5) our model barely reproduces the available constraints on the CO and [CII] luminosity
function of galaxies at $ z \geq 1.5$ and the CO luminosity of
individual galaxies at intermediate redshifts. We argue that this is driven
by a lack of cold gas in galaxies at intermediate redshifts as
predicted by cosmological simulations of galaxy formation. 
\end{abstract}


\section{Introduction}
Our understanding of galaxy formation and evolution has grown a lot
based on the contribution by deep blind fields. These deep fields
mostly focused on the X-ray, optical, near-infrared, sub-mm continuum,
and radio wavelengths. They have contributed tremendously to our
understanding of the star-formation history of our Universe and the
stellar build-up of galaxies and have allowed us to derive a
number of galaxy properties such as stellar masses, star-formation
rates (SFR), morphologies,
and sizes. In particular it has been shown that the
star-formation history of our Universe peaked at redshifts $z\sim2-3$,
after which it dropped to its present day value \citep[e.g.,][for a
recent review
see Madau \& Dickinson 2014]{Madau1996, Hopkins2004, Hopkins2006}. 
\nocite{Madau2014}

The {\it Herschel Space Observatory} \citep{Pilbratt2010} has made significant contributions to our
understanding of galaxy formation and evolution by observing deep
fields of the sub-mm continuum of galaxies \citep[e.g.,][]{Hatlas2010,Hermes2012}.
The Atacama Large (Sub)Millimeter Array (ALMA) allows us to observe deep fields at sub-mm wavelengths with
higher sensitivity and over a larger continuous range of
wavelengths. Additionally, the angular
resolution of interferometers such as ALMA and the Plateau de Bure
Interferometer/Northern Extended Millimeter Array (NOEMA) allows us to pinpoint individual galaxies with much better accuracy compared to single dish observatories
\citep{Decarli2014,Walter2014}. Similar exercises can be carried out
with the Jansky Very Large Array (JVLA) and in the near future the
next generation VLA \citep[ngVLA][]{Carilli2015,Casey2015}. Such efforts can reveal the properties
of atomic and molecular gas in galaxies, a baryonic component not yet
addressed in deep surveys other than local \hi efforts
\citep{HIPASS,ALFALFA}.\footnote{Damped Lyman alpha surveys have
  contributed greatly to our understanding of the global budget of \hi in our
  Universe out to redshifts of $\sim 5-6$ \citep[e.g.,][]{Prochaska2009,Crighton2015}}

Deep surveys are rather expensive, but have advantages
over surveys targeting galaxies based on their stellar masses and/or
SFRs. First of all, blind surveys allow us to detect new classes of
objects previously missed in targeted surveys due to for example
stellar masses and SFRs not fulfilling the selection criteria. More
relevant to this work, blind surveys are ideal to assess the number
densities of different classes of galaxies. With this in mind, blind
surveys with radio and sub-mm instruments are perfectly suited to observe the luminosity
function of the sub-mm continuum of galaxies down to faint
luminosities and high redshifts. Furthermore, due to the high spectral resolution, we
are entering an exciting new era where we can observe the luminosity function
of sub-mm emission lines such as different CO rotational transitions and [CII]. In this paper we make 
predictions for future efforts focusing on the luminosity function of
different CO transitions and [CII] based on a semi-analytic model of
galaxy formation coupled to a radiative transfer code.

Because of its high abundance ($\sim 10^{-4}$ in Milky Way type
galaxies) CO is a bright tracer of the molecular ISM in galaxies. A survey
focusing on CO can therefore effectively trace and provide constraints
on the reservoir of gas
potentially available for star formation (SF) \citep{Walter2014}. Due to its brightness, [CII] is one of the first emission
lines that can be picked up with sub-mm instruments, which makes it a valuable line to find new objects through blind surveys, or
assign spectroscopic redshifts \citep[see for a
review][]{Carilli2013}. In local galaxies [CII] emission correlates
with star-formation \citep{deLooze2011,Herrera2015}, which makes it an extra worthwhile emission
line to go after.

Constraints on the gas content of galaxies are crucial for theoretical models of
galaxy formation. This information is necessary to
break the degeneracies between different physical mechanisms included
in theoretical models such as metal enrichment and feedback
processes. At the same time, models have the potential to provide a
theoretical context for sub-mm emission line deep fields, as this is still an unexplored field.

Recently, theoretical models of galaxy formation started to include recipes to model the
sub-mm line emission from galaxies
\citep[e.g.,][]{Narayanan2008,JP2011,Obreschkow2009COSED,Feldmann2012,Lagos2012,Popping2014radtran,Olsen2015CO,Olsen2015CII}. Semi-analytic
models in particular are are powerful tool to make predictions for CO and [CII] luminosity
functions. Within the semi-analytic framework simplified but
physically motivated recipes are used to track physical processes
such as the cooling of hot gas
into galaxies, star formation, the energy input from supernovae and
active galactic nuclei into the ISM, the sizes of galaxy discs, and the
enrichment of the ISM by supernovae ejecta and stellar winds
\citep[see][for a recent review]{Somerville2014}. The low computational cost of semi-analytic
models makes them a powerful tool to model large volumes on the sky and
provide robust predictions for deep field studies. 

In this work we use an updated version of the
model presented in \citet[P14]{Popping2014radtran}, where we coupled a
radiative transfer model to the \citet[][PST14]{Popping2014sam}
semi-analytical model. The PST14 model has proven to be
successful in reproducing observations of the \hi and \h2 content of galaxies in the
local and high-redshift Universe, such as stellar mass--gas mass
relations, the local \hi and \h2 mass functions, and the sizes of the
gas discs of galaxies. The P14 model successfully reproduces the CO,
[CII], and atomic carbon luminosity of local and high-redshift
galaxies. Updates to the approach presented in P14 concern the coupling between the semi-analytic model and the
radiative transfer code, as well as the sub-grid treatment of
molecular cloud structures. We will present these updates in Section \ref{sec:model}.

This paper is structured as follows. In Section \ref{sec:model} we
present the theoretical model to make predictions for the CO and [CII]
emission of galaxies. We compare model predictions for the scaling
relation between sub-mm lines emission and far-infrared (FIR) luminosity and SFR with
observations of local and high-redshift galaxies in Section
\ref{sec:scaling}. 
We present our predictions for the CO and [CII]
luminosity functions out to $z=6$ in Section
\ref{sec:lumfunc}. We discuss our findings in Section
\ref{sec:discussion} and summarise our work in Section
\ref{sec:summary}. Throughout this paper we adopt a flat $\Lambda$CDM
cosmology with $\Omega_0=0.28$, $\Omega_\Lambda = 0.72$,
$h=H_0/(100\,\rm{km}\,\rm{s}^{-1}\,\rm{Mpc}^{-1}) = 0.7$, $\sigma_8=0.812$, and a
cosmic baryon fraction of $f_b=0.1658$ \citep{Komatsu2009}.

\section{Model description}
\label{sec:model}
\subsection{Galaxy formation model}
\label{sec:sam}
The galaxy formation model used to create a mock sample of galaxies
within a $\Lambda$CDM cosmology was originally presented in
\citet{Somerville1999} and \citet{Somerville2001}. Significant updates
to this model are described in \citet{Somerville2008},
\citet{Somerville2012}, \citet{Porter2014}, PST14, and \citet[SPT15]{Somerville2015}. The model tracks the hierarchical clustering
of dark matter haloes, shock heating and radiative cooling of gas, SN feedback, SF, AGN
feedback (by quasars and radio jets), metal enrichment of the
interstellar and intracluster medium, mergers of galaxies, starbursts,
the evolution of stellar populations, and dust obscuration. The PST14
and SPT15 models include new recipes that track the abundance of
ionised, atomic, and molecular hydrogen and a molecule-based star-formation
recipe. Here we briefly summarise the recipes employed to track the
molecular hydrogen abundance and the molecule-based SF-recipe, as
these set the molecular hydrogen abundance and UV radiation field in
Section \ref{sec:3drealization}. We
point the reader to \citet{Somerville2008}, \citet{Somerville2012},
PST14, and SPT15 for a more detailed description of the model.

To compute the \h2 fraction of the cold gas we use an approach based
on the work by \citet{Gnedin2011}. The authors performed
high-resolution `zoom-in' cosmological simulations with the Adaptive
Refinement Tree (ART) code \citep{Kravtsov99}, including gravity,
hydrodynamics, non-equilibrium chemistry, and simplified 3D
on-the-fly radiative transfer \citep{Gnedin2011}.  The authors find a simple
fitting formula for the \h2 fraction of cold gas based on the
dust-to-gas ratio relative to solar, $D_{\rm MW}$, the ionising background radiation
field, $U_{\rm MW}$, and the surface density of the cold gas,
 $\Sigma_{\rm HI + H2}$. 
The fraction of
molecular hydrogen is given by
\begin{equation}
 f_{H_2} = \left[1+\frac{\tilde{\Sigma}}{\Sigma_{HI+H_2}}\right]^{-2} 
\end{equation}
where
\begin{eqnarray*}
\tilde{\Sigma}  & = &  20\, {\rm M_\odot pc^{-2}} \frac{\Lambda^{4/7}}{D_{\rm MW}} 
\frac{1}{\sqrt{1+U_{\rm MW} D_{\rm MW}^2}}, \\
\Lambda & = & \ln(1+g D_{\rm MW}^{3/7}(U_{\rm MW}/15)^{4/7}),\\
g & = & \frac{1+\alpha s + s^2}{1+s},\\
s &  = & \frac{0.04}{D_*+D_{\rm MW}},\\
\alpha &  = & 5 \frac{U_{\rm MW}/2}{1+(U_{\rm MW}/2)^2},\\
D_* & = & 1.5 \times 10^{-3} \, \ln(1+(3U_{\rm MW})^{1.7}).
\end{eqnarray*}
We assume that the dust-to-gas ratio is proportional to the
metallicity of the gas in solar units $D_{\rm MW} = Z_{\rm gas}/Z_\odot$. We
assume that the local UV background scales with the SFR relative to
the Milky Way value, $U_{\rm MW} = SFR/SFR_{\rm MW}$, where we choose
$SFR_{\rm MW} = 1.0\,\rm{M}_\odot\,\rm{yr}^{-1}$
\citep{Murray2010,Robitaille2010}. Following \citet{Gnedin2011} we take $n_* = 25
\rm{cm}^{-3}$.

We considered other recipes
for the partitioning of \hi and \h2 in PST14 and SPT15. We found that metallicity based recipes that do not
include a dependence on the UV background predict less efficient
formation of \h2, less star-formation, and less metal enrichment at early
times in low-mass haloes ($M_{\rm h}<
10^{10.5}\,\rm{M}_\odot$). PST14 also considered a pressure-based
recipe \citep{Blitz2006}, but found that the pressure-based version
of the model is less successful in reproducing the \hi density of our
Universe at $z>0$.

The SF in the SAM is modelled based on an empirical relationship
between the surface density of molecular hydrogen and the surface
density of star-formation
\citep{Bigiel2008,Genzel2010,Bigiel2012}. Observations of high-density
environments (especially in starbursts and high-redshift objects) have indicated that above
some critical surface density, the relation between molecular hydrogen
surface density and SFR surface density steepens
\citep{Sharon2014,Hodge2014}. To account for this steepening we use the following expression to model star formation 
\begin{equation}
\label{eqn:bigiel2}
\Sigma_{\rm SFR} = A_{\rm SF} \, \Sigma_{\rm H_2}/(10 M_\odot {\rm pc}^{-2}) \left(1+
\frac{\Sigma_{H_2}}{\Sigma_{\rm H_2, crit}}\right)^{N_{\rm SF}},
\end{equation}
where $\Sigma_{\rm H_2}$ is the surface density of molecular hydrogen
and with $A_{\rm
  SF}=5.98\times 10^{-3}\, M_\odot {\rm yr}^{-1} {\rm kpc}^{-2}$,
$\Sigma_{\rm H_2, crit} = 70 M_\odot$ pc$^{-2}$, and $N_{\rm
  SF}=1$.

The sizes of the galaxy discs are important as they set the surface
densities for our \h2 partitioning recipe and SF relation, but will
also control the volume density of the gas when calculating the
line-emission from atoms and molecules. When gas cools onto a galaxy, we assume it initially collapses to form a rotationally supported disc. The scale radius of the disc is computed based on the initial angular momentum of the gas and
the halo profile, assuming that angular momentum is conserved and that
the self-gravity of the collapsing baryons causes contraction of the
matter in the inner part of the halo
\citep{Blumenthal1986,Flores1993,Mo1998}. This approach has shown to
successfully reproduce the evolution of the size-stellar mass relation
of disc-dominated galaxies from $z\sim2$ to $z=0$. PST14 successfully
reproduced the sizes of \hi discs in the local Universe and the
observed sizes of CO discs in local and high-redshift galaxies using
this approach.

We use the approach presented in \citet{Arrigoni2010} to track the
carbon abundance of the ISM. \citet{Arrigoni2010} extended the
Somerville et al. semi-analytic model to include
the detailed metal enrichment by type Ia and type II supernovae and
long-lived stars. With this extension our model tracks the abundances
of 19 individual elements.

FIR luminosities are calculated using the approach presented in
\citet{Somerville2012}. Emission is absorbed by two components. One is
diffuse dust in the disc and the other is associated with the birth
clouds surrounding young star-forming regions. It is then assumed that
all the energy emitted by stars that is absorbed by dust is
re-radiated in the infrared.

\subsection{Creating a 3D realisation of the ISM}
\label{sec:3drealization}
SAMs are a very powerful tool to model the global properties of
galaxies (such as cold gas mass, SFR, stellar mass, and
size). However, they lack detailed information on the spatial distribution of baryons within a
galaxy. In this subsection we
describe the recipes used to create a 3D realisation at parsec-level
resolution of the mock sample of galaxies created by the SAM.

\subsubsection{Gas density}
Under the assumption that cold gas (\hi + \h2) follows an exponential
distribution in the radial and vertical direction, the hydrogen
density at any point in the galaxy at radius $r$ and height $z$ is described as
\begin{equation}
\label{eq:exponential}
n_H(r,z) = n_0(r)\,\exp\left(-\frac{r}{R_g}\right)\exp\left(-\frac{|z|}{z_g(r)}\right),
\end{equation} 
where $n_0(r)$ is the central hydrogen density at any radius $r$, $R_g$ the gas scale length of the galaxy and $z_g(r)$ the
gas scale height. 

The central hydrogen density $n_0(r)$ is given by 
\begin{equation}
  n_0(r) = \frac{M_H}{4\pi m_H\,R^2_g z_g(r)}
\end{equation}
where $M_H$ is the total hydrogen mass (atomic plus molecular) of the
galaxy and $m_H$ the mass of a single
hydrogen atom.

We assume that the gaseous disc is in vertical
equilibrium, where the gravitational force is balanced by the pressure
of the gas. Following \citet{popping2012} and P14 we can then express $z_g(r)$ as
\begin{equation}
z_g(r) = \frac{\sigma_{\rm{gas}}^2}{\pi\,G\,\bigl[\Sigma_{\mathrm{gas}}(r) + 0.1\sqrt{\Sigma_*(r)\Sigma_*^0}\bigr]},
\end{equation}
where $\Sigma_*(r)$ is the stellar surface density, and $\Sigma_*^0$
the central stellar surface density defined as $\frac{M_*}{2\pi r_*^2}$,
with $M_*$ and $r_*$ the stellar mass and scale length of the stellar
disc, respectively. When constructing the gas density profile of the galaxy we adopt a
resolution of 200 pc and integrate the disc out to 8 times its
scale radius. We plot a distribution of the density
weighted average gas density of the modeled galaxies in Figure
\ref{fig:ISMprops}. A more detailed description of the plot will be
given in Section \ref{sec:scaling}.

\subsubsection{\h2 abundance}
The local \h2 abundance of cold gas is dependent on the local cold gas
(column) density, whereas SAMs only provide the global \h2
abundance. The local \h2 abundance is one of the key ingredients when
calculating the level populations of our atoms and molecules of
interest. We therefore calculate the local \h2 abundance in every grid
cell again following the results by \citet{Gnedin2011}. This time the
local \h2 abundance is a
function of gas volume density rather than surface density, together
with the previously defined dust-to-gas ratio relative to solar $D_{\rm MW}$, and
the ionising background radiation field $U_{\rm MW}$ (see Section \ref{sec:sam}). The local fraction of
molecular hydrogen is now given by
\begin{equation}
 f_{H_2} = \frac{1}{1 + \exp{(-4x\,-\,3x^3)}}
\end{equation}
where
\begin{eqnarray*}
x &=& \Lambda^{3/7}\ln{\left(D_{\rm MW}\frac{n_H}{\Lambda
  n_*}\right)},\\
\Lambda & = & \ln{(1+g D_{\rm MW}^{3/7}(U_{\rm MW}/15)^{4/7})},\\
g & = & \frac{1+\alpha s + s^2}{1+s},\\
s &  = & \frac{0.04}{D_*+D_{\rm MW}},\\
\alpha &  = & 5 \frac{U_{\rm MW}/2}{1+(U_{\rm MW}/2)^2},\\
D_* & = & 1.5 \times 10^{-3} \, \ln(1+(3U_{\rm MW})^{1.7}).
\end{eqnarray*}

Following \citet{Gnedin2011} we take $n_* = 25\,
\rm{cm}^{-3}$. We normalize the sum of the local \h2 masses to the global \h2 mass to
assure that the global \h2 mass is conserved.

\subsubsection{Radiation field}
We derive the FUV (6--13.6 eV) field strength, $G_{\rm UV}$, by relating the SFR
density to the FUV-radiation field as
\begin{equation}
\frac{G_{UV}}{G_0}  = \frac{\rho_{\rm SFR}}{\rho^0_{\rm SFR}},
\end{equation}
where $\rho_{\rm SFR}$ is the density of SF in
$M_\odot\,\rm{yr}^{-1}\,\rm{kpc}^{-3}$, $\rho^0_{\rm SFR}$ is the
average SFR density in the MW, and $G_0 =
1.6\times10^{-3}\,\rm{erg}\,\rm{cm}^{-2}\,\rm{s}^{-1}$ (the Habing
Flux). We scale the density of SF as a function of
the molecular hydrogen density in every grid cell with $\rho_{\rm
  SFR}=\rho_{\rm H2}^{1.5}$, normalising it such
that the total integrated SFR equals the SFR as predicted by our SAM. We take $\rho^0_{\rm SFR} = 0.0024\,
M_\odot\,\rm{yr}^{-1}\,\rm{kpc}^{-3}$ \citep{Olsen2015CO}, which
corresponds to the SFR density in the central 10 kpc of our MW. The distribution of the density
weighted average UV radiation field in the modeled galaxies is shown in Figure
\ref{fig:ISMprops}.

\subsubsection{Abundances}
The CO abundance of the cold gas is calculated as the amount of carbon locked up in
CO.  The fraction of the carbon mass locked up in
CO has an explicit dependence on metallicity. Following  \citet{Wolfire2010} we
calculate this fraction as 
\begin{multline}
f_{\rm CO} = f_{\rm H2} \times \\
e^{-4\,\big(0.52 -
0.045\,\ln{\frac{G_{UV}/(1.7G_0)}{n_H}} -
  0.097\ln{\frac{Z_{\rm gas}}{Z_\odot}}\big)/A_V},
\end{multline}
where $A_v=n_H
R_\mathrm{grid}(Z_\mathrm{gas}/Z_\odot)/1.87\times10^{21}$ mag,
with $R_{\rm grid}$ the size of a grid cell in cm.

The remaining carbon is either ionised or atomic. We assume that the
atomic and ionised carbon are equally distributed at $A_V = 1$ mag. At
$A_V=0$ mag all the carbon is ionised, whereas at $A_V = 10$ mag only 10\% of the
carbon is ionised. These numbers reach good agreement with predictions from \citet{Meijerink2005} for the typical range of densities and radiation fields relevant to our work. We perform logarithmic interpolation between these points to calculate
the abundance of atomic and ionised carbon at any $A_V$.

\subsubsection{Temperature}
We calculate the temperature of the gas and dust using the {\sc DESPOTIC} package
\citep{Krumholz2013}. Unless stated otherwise the physical
parameters match the defaults in {\sc DESPOTIC}.

The temperature of the cold gas and dust is set by a balance of heating and
cooling processes. Heating terms that are included are cosmic ray
heating, photo-electric heating, gravitational heating, and the
exchange of energy between dust and gas. The primary cooling
mechanism for the gas is line radiation. We take the cooling through
CO, atomic carbon [C], and ionised carbon C$^+$ into account. We refer
the reader to \citet{Krumholz2013} for a detailed explanation of the
different heating and cooling terms. We set the temperature of the
cosmic-microwave background at the redshift of the galaxy as a lower
limit on the gas and dust temperature. We note that the adopted approach
for calculating temperature is a significant improvement with respect to
P14, where a simplified model was assumed only including the cooling
though oxygen and ionised carbon. The addition of CO cooling in the
densest environments allows for lower temperatures, which additionally
suppresses the amount of CO emission. We plot a distribution of the density
weighted average gas and dust temperatures of the galaxies in Figure \ref{fig:ISMprops}.

\subsubsection{Velocity field and turbulence}
To trace the absorption of photons along the line of sight 
information about the velocity field of the galaxy is necessary. We
derive the velocity field following the approach presented P14, where the radial velocity
profile of a galaxy is constructed based on a component from the
bulge, disc, and halo, respectively. We assume a vertical velocity
dispersion $\sigma_{\rm gas}$ of $10\,\rm{km}\,\rm{s}^{-1}$
\citep{Leroy2008}. The local non-thermal turbulent velocity dispersion
is derived as the standard deviation of the velocities in
the nearest neighbouring cells in all directions. 

\subsection{Radiative-transfer and line tracing}
\label{sec:radtran}
We use an updated version of the advanced fully three-dimensional radiative-transfer code
$\beta$3D \citep{Poelman2005,Poelman2006}, optimised for heavy memory
usage by \citet{JP2011}. To calculate the level populations of the
molecule or atom of interest, $\beta$3D takes the escape
probability of photons out of a molecular cloud along 6 directions
into account. The optimised version was initially developed
to calculate the three-dimensional transfer of line radiation in 256 x
256 x 128 element data cubes at a spatial resolution of 0.25 pc. P14
optimised this code to calculate the line properties of galaxy sized
objects with much lower spatial resolution. 

Calculating the emitted radiation from an atomic or molecular species
requires solving for the number density of atoms or molecules in the level
of interest. It also requires calculating the probability that a photon at
some position in the cloud can escape the system. The basic assumption
in the radiative transfer calculation is that the levels of the atomic or molecular species are in statistical
equilibrium. This implies that the rate of transitions out of each level is balanced by the rate of transitions
into that level. For a multi-level molecule, this can be expressed
using the equations of statistical equilibrium for each bound level
$i$, with population density $n_i$, and energy $E_i$, as 
\begin{equation}
n_i\sum_j R_{ij} = \sum_j n_j R_{ji},
\end{equation}
where the sums are over all other bound levels $j$. $R_{ij}$ gives the rate at which transitions from level
$i$ to $j$ occur. These equations are supplemented by the constraint
that the sum of all populations $n_{i}$ equals the density of the
atomic or molecular species $x$ in all levels,
\begin{equation}
n_{x} = \sum_j n_j,
\end{equation}
and together these equations constitute a complete system that can be
solved iteratively.

The rates $R_{ij}$ are expressible in terms of
the Einstein $A_{ij}$ and $B_{ij}$ coefficients, and the collisional
excitation $(i<j)$ and de-excitation $(i>j)$ rates $C_{ij}$:
\begin{equation}
R_{ij} = \left\{
\begin{matrix}
A_{ij} + B_{ij}\langle J_{ij} \rangle + C_{ij}, & \quad E_i > E_j,\\
&\\
B_{ij}\langle J_{ij} \rangle + C_{ij}, & \quad E_i < E_j.
\end{matrix}\right.
\end{equation}
The Einstein $A_{ij}$ coefficient gives the rate of an electron decaying
radiatively from an upper state $i$ to a lower state $j$. The Einstein $B_{ij}$
rate gives the rate of an atom or molecule absorbing a photon, which causes
an electron to be excited from a lower state $j$ to an upper state $i$. The collision rate $C_{ij}$ sets the coupling between the
excitation of the atom or molecule and the kinetic energy of the gas and
depends (for each collisional partner such as atomic and molecular
hydrogen and helium) on the kinetic temperature of
the gas. $\langle J_{ij} \rangle$ is the mean integrated radiation
field over $4\pi$ steradian at a
frequency $\nu_{ij}$ corresponding to a transition from level $i$ to
$j$ and is given by 
\begin{equation}
\langle J_{ij} \rangle = (1 - \beta_{ij}) S_{ij} + \beta_{ij}B_{ij}(\nu_{ij}),
\end{equation}
where $\beta_{ij}$ is the escape probability of a photon and $S_{ij}$
is the source function. The background radiation
$B_{ij}(\nu_{ij})$ comes from the infrared emission of
dust at a temperature $T_d$ and the temperature of the Cosmic
Microwave Background (CMB) $T_{\rm CMB}$ at the
redshift of interest. The background radiation field is given by
\begin{equation}
B_{ij}(\nu_{ij}) = B(\nu_{ij},T = T_{\rm CMB}) + \tau_d(\nu_{ij})B(\nu_{ij},T_d),
\end{equation}
where $\tau_d(\nu_{ij}) = \tau_{100\mu m}(100\,\mu m/\lambda)$. We adopt
a value of $\tau_{100\mu m} = 0.001$ \citep{Hollenbach1991}.

The source function is defined as the ratio between the emission
coefficient and the absorption coefficient. It is a measure of how
photons in a light beam are absorbed and replaced by new emitted photons by the system it passes through and is given by 
\begin{equation}
S_{ij} = \frac{n_j A_{ij}}{n_i B_{ij} - n_j B_{ji}} =
\frac{2h\nu_{ij}^3}{c^2}\left[\frac{n_j g_i}{n_i g_j} - 1\right]^{-1},
\end{equation}
where $g_i$ and $g_j$ are the statistical weights of level $i$ and
$j$, $n_i$ and $n_j$ the population density in the $i$th and $j$th
level, $h\nu_{ij}$ is the energy difference between the levels $i$
and $j$, and $c$ the speed of light. 

As mentioned above, calculating the emitted intensity by a molecular
cloud requires knowledge of the escape probability of the emitted
photons. For a sphere, the probability of a photon emitted in the transition
from level $i$ to level $j$ to escape the cloud is given by 
\begin{equation}
\beta_{ij} = \frac{1 - \exp(-\tau_{ij})}{\tau_{ij}},
\end{equation}
where $\tau_{ij}$ is the optical depth in the line. The optical depth
in the line over a distance running from $s_1$ to $s_2$ is given by 
\begin{equation}
\tau_{ij} =
\frac{A_{ij}c^3}{8\pi\nu_{ij}^3}\int\limits_{s_{1}}^{s_{2}}
\frac{n_i}{\Delta v_d}\left[ \frac{n_j g_i}{n_i g_j}-1\right] \rm{d}s,
\end{equation}
where $\Delta v_d$ is the velocity dispersion of the gas due to local
turbulence in the cloud. 

The emerging specific intensity from a single molecular cloud can now be
expressed as
\begin{equation}
dI^z_\nu =
\frac{1}{4\pi}n_iA_{ij}h\nu_{ij}\beta(\tau_{ij})\,\bigl(\frac{S_{ij}- B_{ij}(\nu_{ij})}{S_{ij}}\bigr)\phi(\nu)dz,
\end{equation}
where $dI^z_\nu$ has units of erg cm$^{-2}$ s$^{-1}$ sr$^{-1}$
Hz$^{-1}$, $\phi(\nu)$ is the profile function, which is the Doppler correction to the photon frequency due to
local turbulence inside the cloud and large scale bulk
motions, and  $B_{ij}(\nu_{ij})$ is the local continuum background radiation at
the frequency $\nu_{ij}$.

The sizes of individual molecular clouds in
  galaxies are often much smaller than the 200 pc resolution of our grid. To account for this we assume that a grid cell is made up by small molecular
clouds all with a size of the Jeans length that belongs to the typical
temperature and density of the grid cell.

To include the effects of clumping in a molecular cloud we multiply
the collisional rates $C_{ij}$ with a clumping factor
$f_{\rm cl}$ \citep{Krumholz2013}, the factor by which
the mass-weighted mean density exceeds the volume-weighted mean
density. In a supersonic turbulent medium this factor can be
approximated by 
\begin{equation}
f_{\rm cl} = \sqrt{1 + 0.75 \Delta v_d^2/c^2_{\rm s}},
\end{equation}
where $c_{\rm s}$ is the sound speed of the medium \citep[e.g.,][]{Ostriker2001,Lemaster2008, Federrath2008,Price2011}.

We assign the individual
molecular clouds a relative velocity with respect to each other drawn from a gaussian
distribution centered around 0 $\rm{km}\,\rm{s}^{-1}$, with the
velocity dispersion determined for that grid cell as the
standard deviation. We calculate the
contribution of each of these individual molecular clouds within the
sub-grid to the emitted radiation and take the overlap in optical depth
space of the molecular clouds into account. This is a
  fundamental update to the sub-grid treatment of the radiative
  transfer approach with respect to P14, where the individual
  molecular clouds had the same relative velocity. We expect the
  optical depth within a grid cell to be smaller than in P14, effectively allowing
  more emission to escape from dense regions.

The line intensity
escaping the galaxy is computed using a ray-tracing approach,
including the effects of kinematic structures in the gas and optical
depth effects along the line-of-sight towards the observer. The emerging specific intensity
is dependent on the escape probabilities within a grid cell
as well as connecting adjacent grid points along the line of
sight. This makes our approach more physical compared to the purely
local nature of the LVG approximation \citep[e.g.,][]{Weiss2005}. 

Level populations for $^{12}$CO and C$^+$ are calculated using rate coefficients available in the LAMDA database
\citep{Schoier2005}. We use \h2 and helium as the main collision partners for the
radiative transfer calculations for CO. The collisional partners for
ionised carbon are \h2, \hi, and an
electron abundance that scales with the C$^+$ abundance. The densities of the collisional partners are derived from
the galaxy formation model described in Section \ref{sec:sam}.

\begin{figure}
  \includegraphics[width = 1.0\hsize]{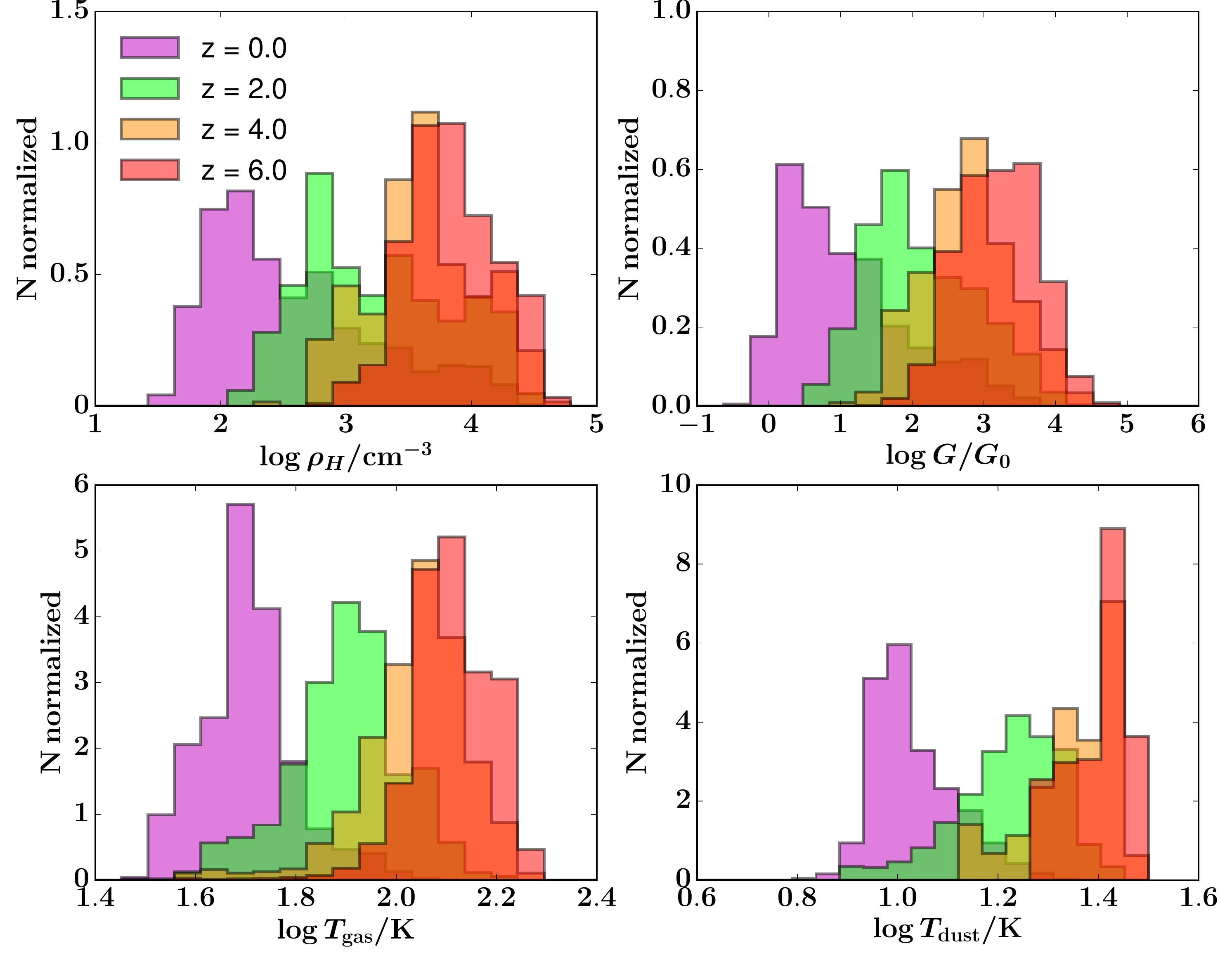}
  \caption{ The distribution of the density weighted average gas density
    (top left), UV radiation field (top right), gas temperature
    (bottom left), and dust temperature (bottom right) for central
    star-forming galaxies at
    redshifts $z=0,\,2,\,4$ and $6$.\label{fig:ISMprops}}
\end{figure}

\begin{figure*}
  \includegraphics[width = 1.0\hsize]{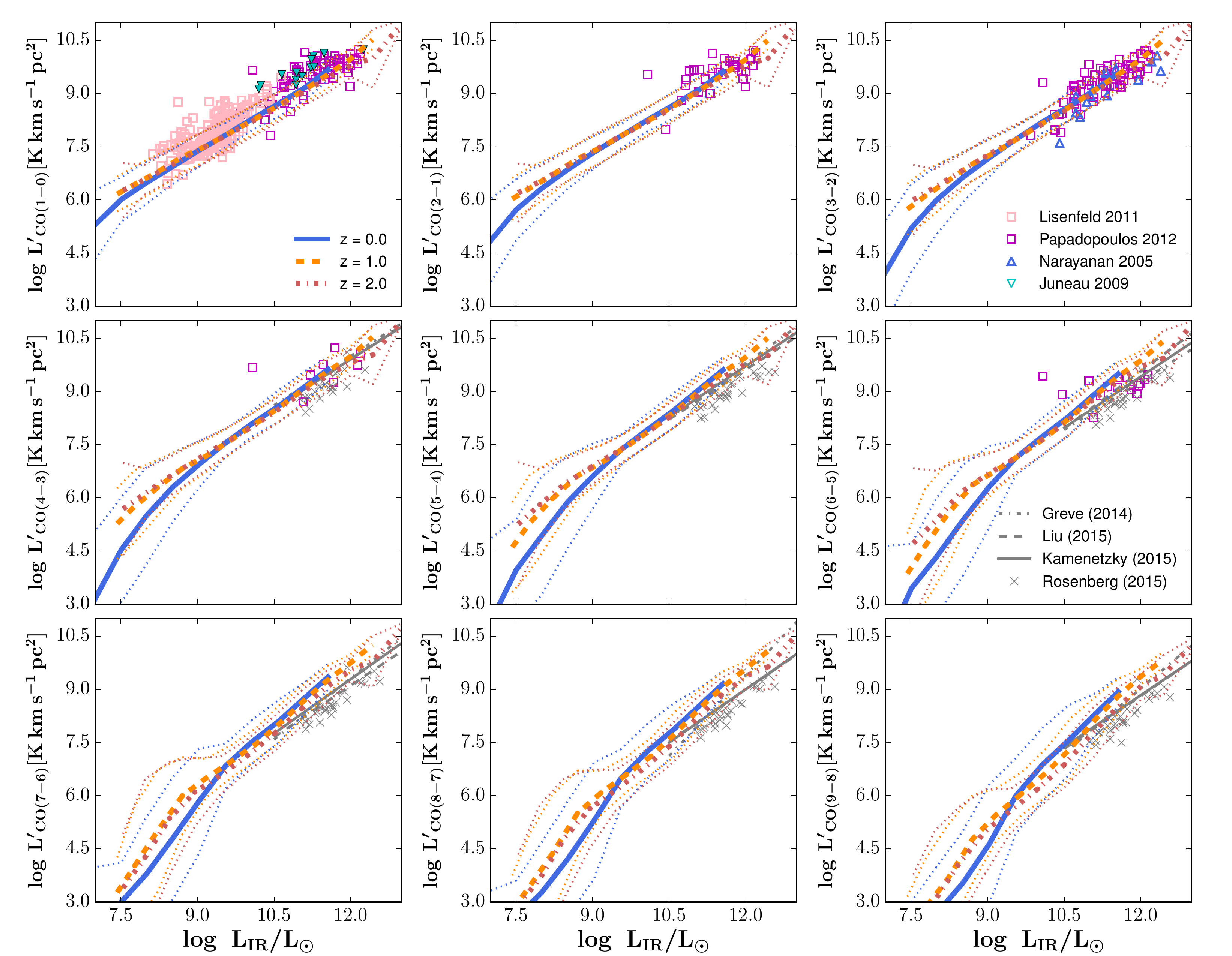}
  \caption{CO line-luminosity of CO J$=$1-0 up to CO J$=$9-8 as a
    function of FIR luminosity at redshifts $z=0$, $z=1$, and $z=2$. Model results are compared to
    observations taken from \citet{Narayanan2005}, \citet{Juneau2009}, \citet{Lisenfeld2011}, \citet{Papadopoulos2012},
    \citet{Greve2014}, \citet{Liu2015}, \citet{Rosenberg2015}, and \citet{Kamenetzky2016}. The thick
    lines show the median of the model predictions, whereas the dotted
    lines represent the two sigma deviation from the median. 
\label{fig:scaling_FIR}}
\end{figure*}

\begin{figure*}
 \includegraphics[width = \hsize]{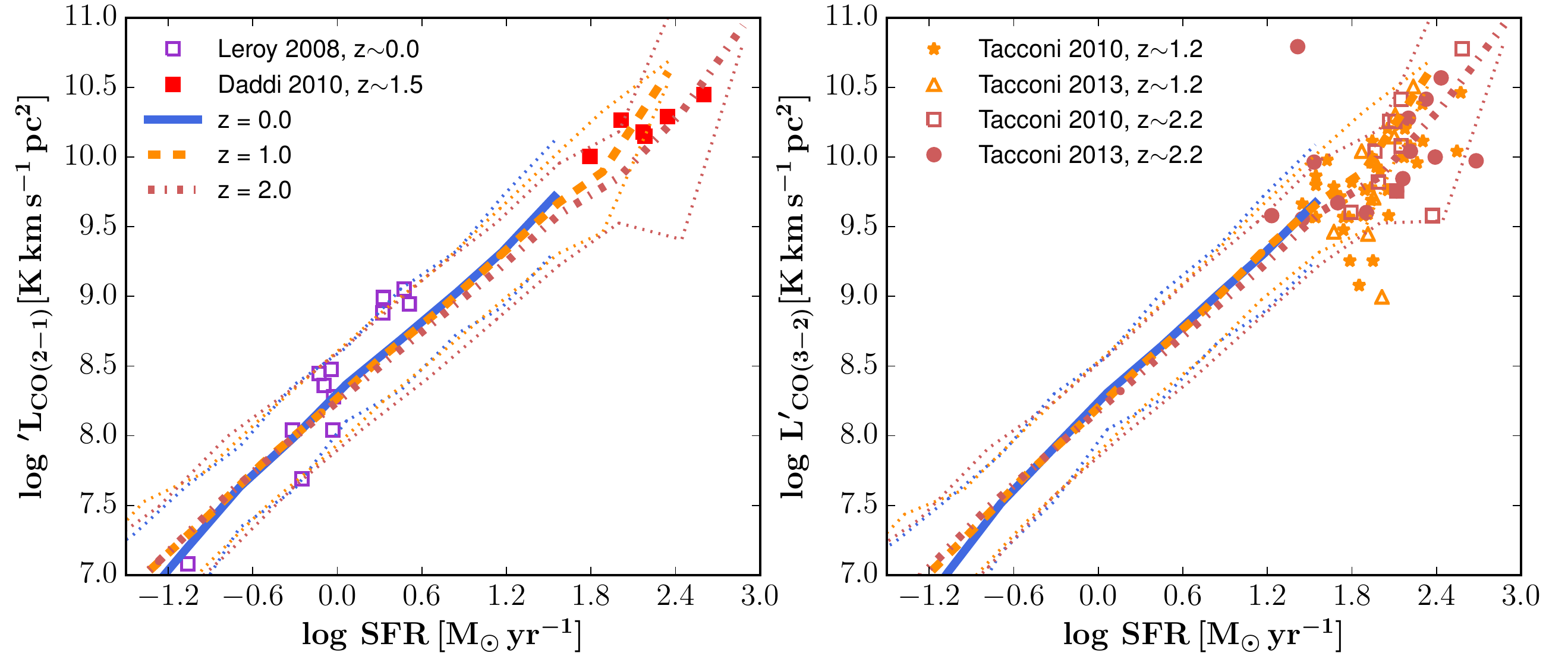}
 \caption{CO line luminosity of CO J$=$2-1 (left panel) and CO J$=$3-2
   (right panel) as a function of SFR for modeled galaxies at $z=0$,
   $z=1$, and $z=2$. Observations are taken from \citet{Leroy2009CO},
   \citet{Daddi2010}, \citet{Tacconi2010}, and \citet{Tacconi2013}. \label{fig:SFR_CO}}
 \end{figure*}

\section{CO and [CII] scaling relations}
\label{sec:scaling}
In this section we present our model predictions for the CO and [CII] line
luminosities of galaxies as a function of SFR and IR
luminosity. Very similar predictions were shown in P14. In this work we have significantly updated the
recipes for the cooling of gas and the sub-grid treatment of the
radiative transfer approach. We therefore believe it is good to reassure
ourselves that our model reaches good agreement with
observations. Furthermore, in this work we extend the comparison
between model and observations out to CO J$=$9--8.

The simulations were run on a grid of haloes with viral
masses ranging from $10^9$ up to $5 \times 10^{14}\,M_\odot$, with a
resolution down to $10^7\,M_\odot$. From these simulations we selected all
central galaxies with a molecular hydrogen gas mass more massive than
the mass resolution of our simulations. 
In this Section we restricted our analysis to
central star forming galaxies, selected using the criterion $\rm{sSFR} >
1/(3t_H(z))$, where $\rm{sSFR}$ is the galaxy specific star-formation
rate and $t_H(z)$ the Hubble time at the galaxy's redshift. This
approach selects similar galaxies to commonly used observational
methods for selecting star-forming galaxies \citep[e.g.,][]{Lang2014}.

Before presenting actual scaling relations we show normalized 
  distribution functions of the density weighted average gas density,
  UV radiation field, gas temperature and dust temperature in the
  modeled central star-forming galaxies in Figure \ref{fig:ISMprops}. This should give the reader a
  sense of the evolving ISM in the modeled galaxies. We find that the density and radiation
  field of the ISM in galaxies decrease with cosmic time. Similarly,
  the average temperatures of dust and gas also decrease with cosmic
  time. The contribution of the CMB to the temperature of the dust is visible in
the lower limit of dust temperatures at $z=4$ and $z=6$. We note that
these properties are density weighted averages and can vary between the grid cells within a galaxy.

\subsection{CO}
In Figure \ref{fig:scaling_FIR} we show the predicted CO J$=$ 1--0 out to CO
J$=$9--8 line luminosity of galaxies at redshifts z$=$0, 1, and 2 as a
function of their FIR luminosity. Our model
predictions at z$=$0 are in good agreement with observational
constraints from CO J$=$1--0 out to CO J$=$5--4. We compare our
predictions for the CO J$=$4--3 and higher line emission of galaxies
with data and fits from the literature
\cite{Greve2014,Liu2015,Rosenberg2015,Kamenetzky2016}.  
% The authors made a
% compilation of high-J CO observations of galaxies and star-forming
% regions in local galaxies, and found that a single
% powerlaw can describe the relation between FIR luminosity and CO
% luminosity. We note, however, that below FIR luminosities of
% $\sim10^{10}\,\rm{L}_\odot$ the fit is driven by individual
% star-forming regions, rather than global galaxy luminosities such as predicted by our model. 
Our model results
are in good agreement with the observations for CO
J$=$4--3 and CO J$=$5--4. We predict slightly too much line emission
for the higher CO rotational transitions at z$=$0 compared with observations. In P14 we predicted too much CO emission for CO J$=$2--1 and
higher rotational transitions. Overall we find that the agreement between
our model and the $z=0$ observations has improved compared to the P14
results.

We find hardly any time evolution in the relation between FIR luminosity
and CO line luminosity for galaxies with FIR luminosities fainter
than $\sim10^{11.5}\,\rm{L}_\odot$. We find minor evolution towards
FIR-brighter galaxies, where the CO luminosity of galaxies decreases
with increasing redshifts. This supports a redshift independent relation
between the FIR and CO line luminosity of galaxies. Similar to our $z=0$ predictions, we predict slightly too
much line emission for CO J$=$6--5 and higher rotational
transitions.

In Figure \ref{fig:SFR_CO} we plot the predicted CO J$=$2--1 and CO J$=$3--2 line
emission of galaxies as a function of galaxy SFR  at
redshift $z=0, 1$ and 2. We compare our
predictions with direct observations of the CO emission lines taken
from \citet{Leroy2009CO}, \citet{Daddi2010}, \citet{Tacconi2010}, and
\citet{Tacconi2013}. We reach good agreement with the observed CO
luminosities at all redshifts. 

We find mild
evolution in the relation between CO luminosity and SFR towards the
galaxies with highest SFRs (SFR$> 15\,\rm{M}_\odot\,\rm{yr}^{-1}$), where the CO
luminosity of galaxies slightly decreases towards lower redshifts. The rate of this evolution is less than we found in P14. We ascribe that to
a better treatment of the sub-grid physics introduced to properly
account for optical depth effects within a grid cell. In P14 we
did not introduce a local velocity dispersion between the individual
molecular clouds in a grid cell. This led to optical
depths that were slightly too large, which resulted in an
underestimate of the emitted CO radiation in dense (high-redshift)
objects. The lack of galaxies at $z=0$ with SFR
$\sim100\,\rm{M}_\odot\,\rm{yr}^{-1}$ is because of the quenching of
actively star-forming objects.

Overall we find that our model is able to reproduce available
observations of CO line luminosities very well out to transitions of
CO J$=$5-4. We predict slightly too much emission towards the highest
transitions we explored. In the remainder of this paper
we will focus on CO line transitions ranging from CO J$=$1-0 to CO J$=$6-5.
\begin{figure}
\includegraphics[width = \hsize]{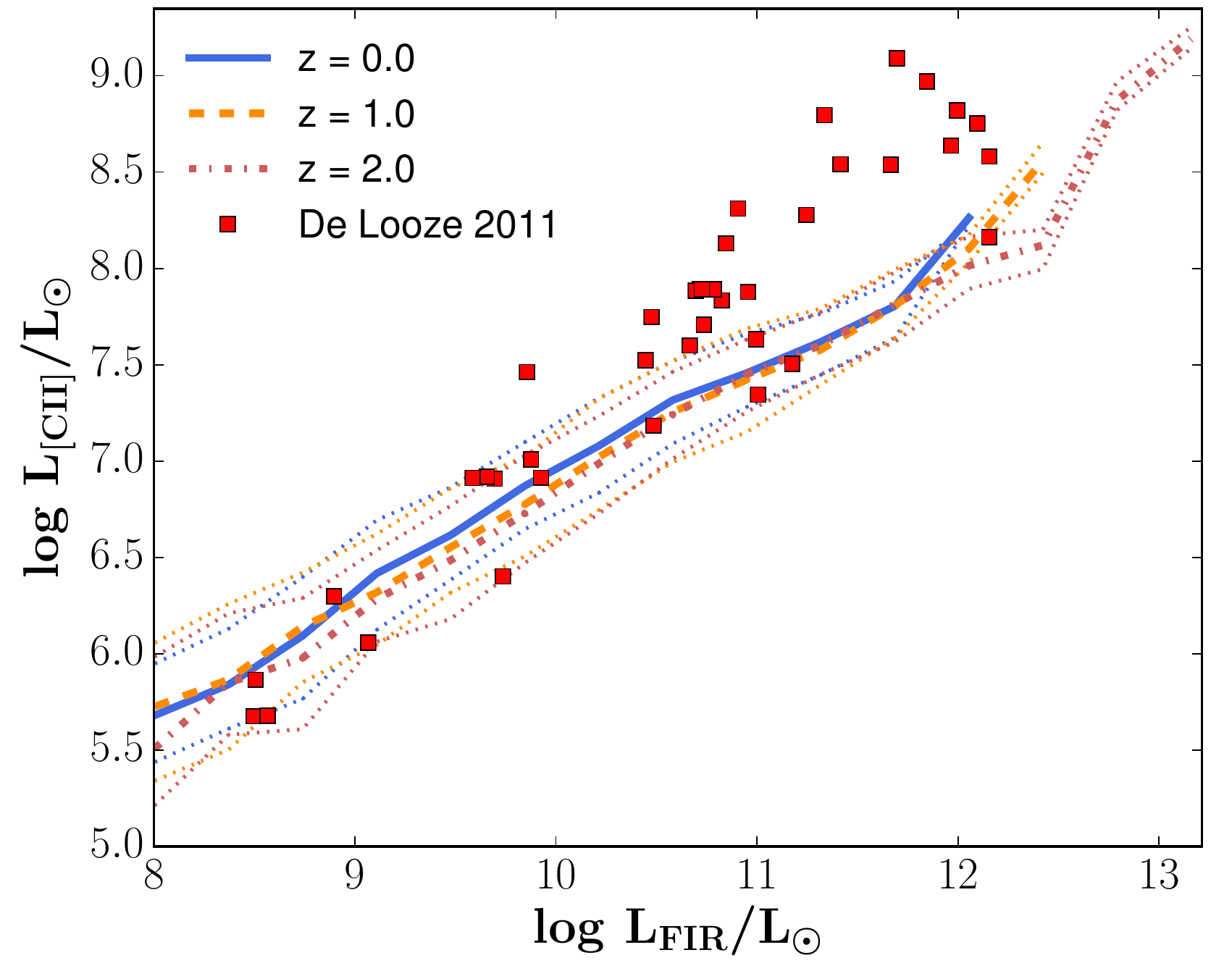}
\caption{Luminosity of the atomic cooling line [CII] (158 $\mu$m) as a function of FIR luminosity for galaxies
  at $z=0$, $z=1$, and $z=2$. Observations at $z=0$ are from \citet{deLooze2011}.\label{fig:cII_scale}}
\end{figure}

\begin{figure*}
 \includegraphics[width =\hsize]{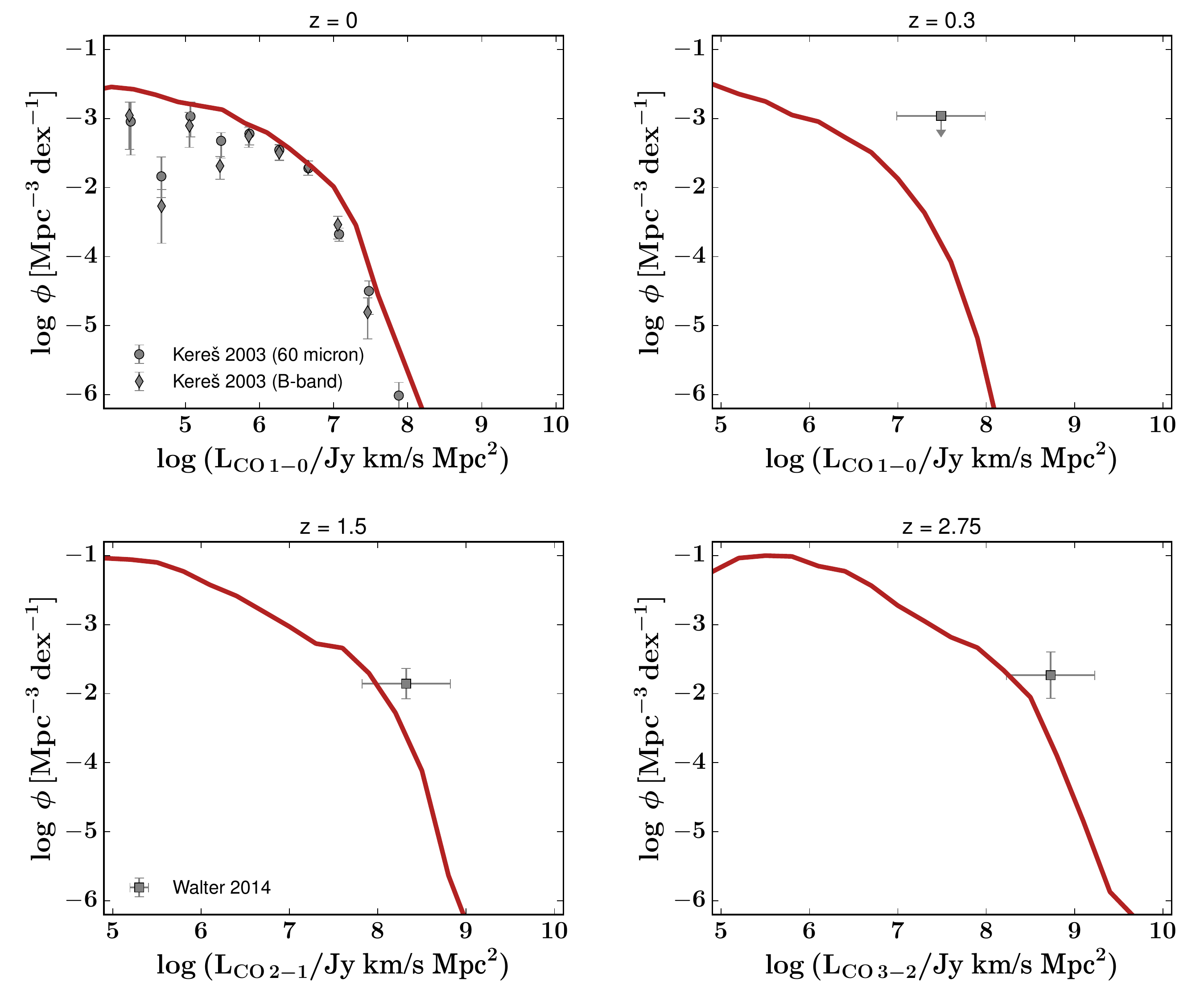}
 \caption{Comparison between model predictions and observations of the
   CO luminosity at $z=0.0$ (CO J$=$1--0;
   upper left panel), $z=0.3$ (CO J$=$1--0; upper right panel), $z=1.5$
   (CO J$=$2-1; lower left panel), and $z=2.75$ (CO J$=$3-2; lower
   right panel). Predictions are compared to observational
   constraints from \citet{Keres2003} and \citet{Walter2014}.
 \label{fig:CO_literature}}
 \end{figure*}

 \begin{figure*}
 \includegraphics[width = \hsize]{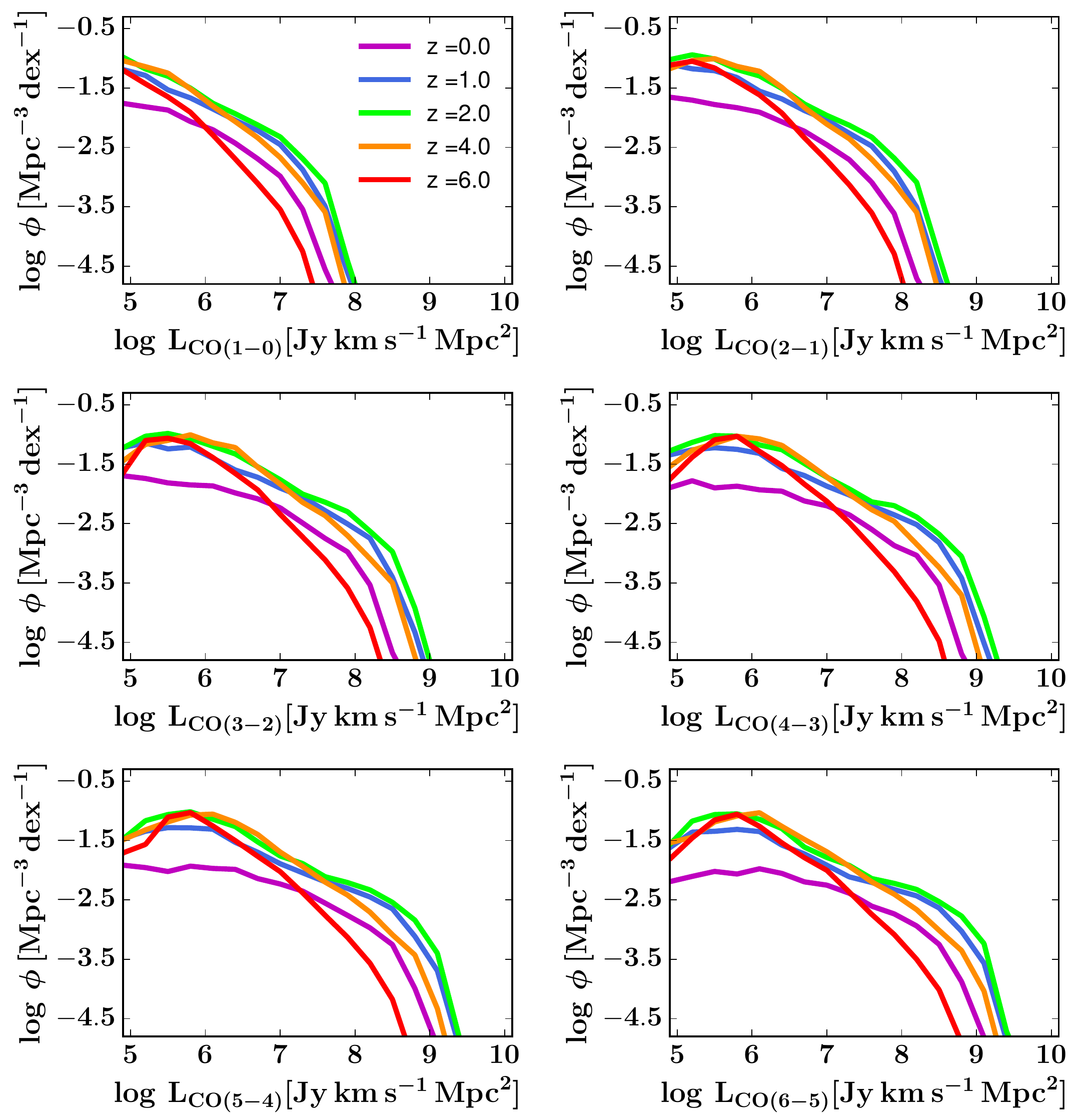}
 \caption{Model predictions of the CO J$=$1--0 up to the CO J$=$6--5
   luminosity function of galaxies from $z=0$ out to $z=6$.
 \label{fig:CO_evol}}
 \end{figure*}

 \begin{figure*}
 \includegraphics[width = 1.0\hsize]{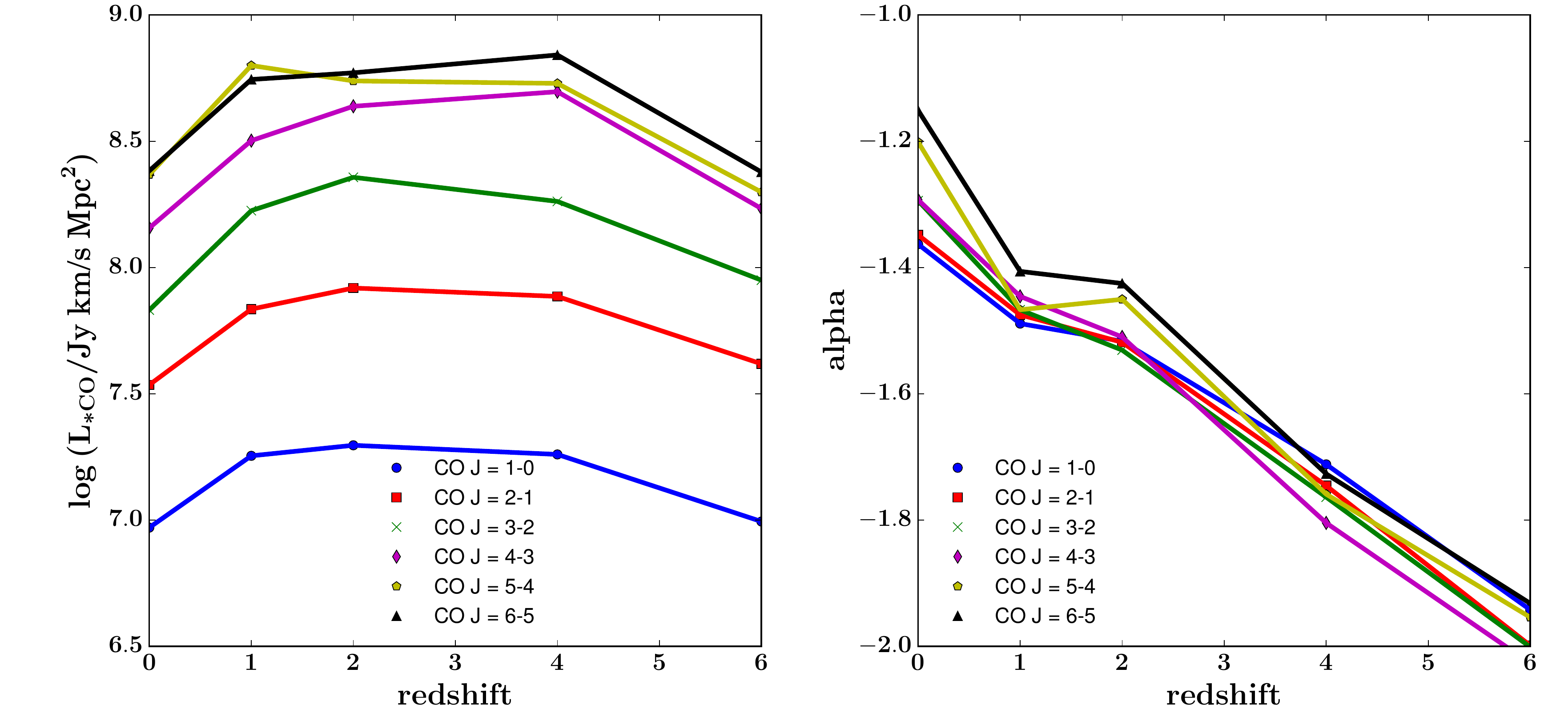}
 \caption{Evolution of the turning point of the Schechter function
   $L_*$ (left) and the slope of the powerlaw component of the Schechter function
   $\alpha$ (right) for predicted CO luminosity function from CO J$=$1--0 up to CO
   J$=$6--5 from redshift $z=6$ to $z=0$. \label{fig:Lstar_evol}}
 \end{figure*}

 \begin{figure*}
 \includegraphics[width = 1.0\hsize]{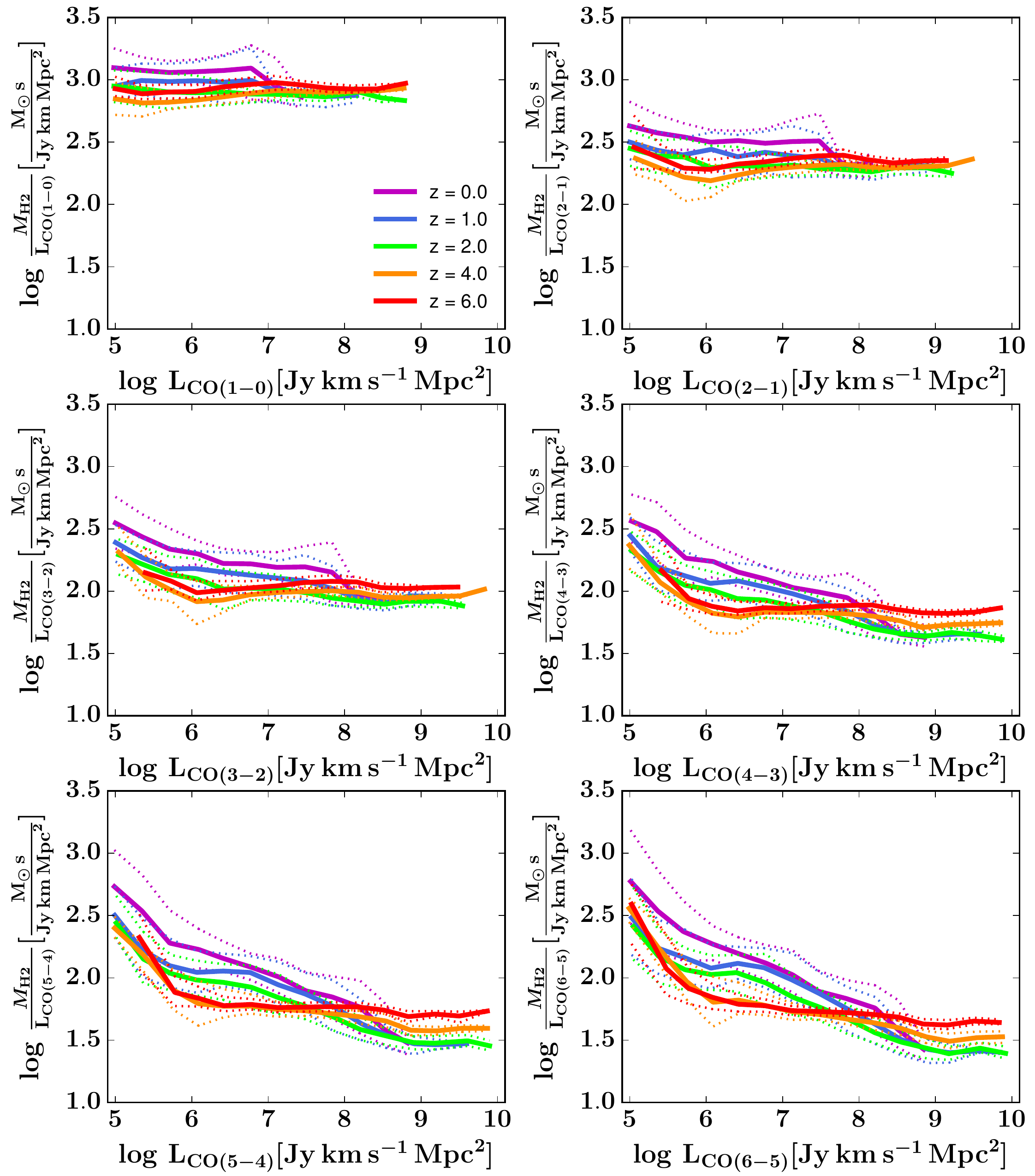}
 \caption{ The ratio between \h2 mass and CO luminosity for CO J$=$1--0
   up to CO J$=$6--5 at redshifts $z=0$ to
   $z=6$. The solid lines show the median of the model predictions,
   whereas the dotted lines represent the two sigma deviation from the
   median. Note the increase in the \h2-to-CO ratio at low
   luminosities with increasing redshift, especially for the higher
   rotational CO transitions.\label{fig:LCO_alpha}}
 \end{figure*}

\subsection{[CII]}
We plot the [CII] luminosity of galaxies as a function of
their FIR luminosity in Figure \ref{fig:cII_scale}. We find decent
agreement between our model
predictions at $z=0$ and the \citet{deLooze2011} observations of the
[CII] luminosity of galaxies at FIR luminosities less than
$10^{11}\, L_\odot$.We underpredict the [CII]
luminosity of FIR-brighter galaxies. We
find hardly any evolution in the [CII] luminosity of galaxies at fixed
FIR luminosity from $z=0$ to $z=2$.

\section{CO and [CII] luminosity functions}
\label{sec:lumfunc}
In this section we present our predictions for the CO
and [CII] luminosity function of galaxies at different
redshifts. Unlike in Section \ref{sec:scaling} we selected
all galaxies (both centrals and satellites and selection criteria
based on the SFR of galaxies was applied). We first compare our predictions with
observational estimates of the CO luminosity function from the
literature at different redshifts. We then present predictions for future observations,
and focus on the evolution in the shape of the CO luminosity
function. We finish by presenting
the evolution of the [CII] luminosity function and its shape. 

We plot the different CO luminosity functions
in terms of the velocity integrated luminosity $\rm{L}_{CO}$ with units of
$\rm{Jy}\,\rm{km}\,\rm{s}^{-1}\,\rm{Mpc}^2$, as this gives a better
representation through which of the CO J-transitions the dominant
CO cooling occurs. These units can easily be converted into 
commonly used brightness temperature luminosities $\rm{L'}_{CO}$ in $\rm{K}\,\rm{km}\,\rm{s}^{-1}\,\rm{pc}^2$ using the equation
\begin{equation}
\rm{L'}_{CO} = \frac{c^3}{8\pi k_B\nu^3_{\rm rest}}\rm{L}_{CO},
\end{equation}
where $k_B$ is the Boltzmann constant and $\nu_{\rm rest}$ the
rest frequency (i.e. not redshifted) of the transition.

\subsection{Carbon Monoxide}
\subsubsection{Comparison with the literature}
Fig. \ref{fig:CO_literature} shows a comparison between predicted CO
luminosity functions and observational constraints from
\citet{Keres2003} at $z=0.0$ and \citet{Walter2014} at $z=0.3,\, z=1.5$,
and $z=2.75$. To avoid including an additional uncertainty in the conversion of high CO J-transitions to
CO J$=$1--0, we chose to carry out the comparison for the CO
J-transitions that were originally observed

Our model predictions for CO J$=$1--0 at $z=0$ are in good agreement
with the observed CO luminosity function by \citet{Keres2003}. We
slightly overpredict the number of galaxies with CO J$=$1--0 luminosities
less than $10^5\,\rm{Jy}\,\rm{km}\,\rm{s}^{-1}\,\rm{Mpc}^2$, and
properly reproduce the number of galaxies with brighter CO
luminosities. Our model predictions fall within the uncertainty
regions of observational constraints on the CO J$=$1--0, J$=$2--1, and J$=$3--2,
luminosity function of galaxies at redshifts $z=0.3,$ $z=1.5$, and
$z=2.75$, respectively. It must be said that our predictions at
$z=1.5$ and especially $z=2.75$ are only barely in agreement with the available
observational constraints. \citet{Walter2014} showed that other models
fail to reproduce their observational constraint at
$z=2.75$. We elaborate further on this in Section
\ref{sec:discussion}. Unfortunately, there are currently no direct
constraints available for the low-mass end of the CO luminosity
function at $z>0$.

\subsubsection{Evolution of the CO luminosity function}
In this section we present our predictions for the evolution of CO
luminosity functions. It is expected that in the near
future more and more deep blind fields and indirect efforts will provide 
constraints on the CO luminosity function, ultimately probing the
molecular hydrogen density of our Universe \citep[through an CO-to-\h2 mass
conversion factor, see][for a review]{Bolatto2013} 

Fig. \ref{fig:CO_evol} shows a clear pattern in the evolution of the luminosity
function of different CO J-transitions with time. The number density
of galaxies increases
from $z=6$ to $z=4$ after which the number density stays
remarkably constant till $z=1$. This behaviour holds over the entire
luminosity range probed. At $z < 1.0$ the number density decreases
over the entire range of CO luminosities, independent of the
CO J-transition. This type of evolution (a relatively constant
luminosity function at redshifts $z=1-4$ and decreasing number
densities at later times) was also seen in the model predictions for
the \h2 mass function of galaxies (PST14).

\subsubsection{Shape of the CO luminosity function}
Since the
different CO J-transitions trace different phases of the molecular
ISM, differences in evolution may hint towards differences in the
composition of the ISM in galaxies with time. To better quantify the
evolution we fit our predicted CO luminosity functions with a Schechter
function
\begin{equation}
\phi(L_{\rm CO}) = \frac{dn}{d\log{L_{\rm CO}}} = \ln{10}\,\phi_* \,
\bigl(\frac{L_{\rm CO}}{L_*}\bigr)^{\alpha + 1}\, e^{-\frac{L_{\rm CO}}{L_*}}.
\end{equation}
In this equation $L_*$ is the luminosity at which the Schechter
function turns from a powerlaw into an exponential, $\alpha$ is the
slope of the powerlaw, and $\phi_*$ is the normalisation of the
luminosity function. In the remainder of this work we will focus on the turning
point $L_*$ and the slope of the powerlaw component $\alpha$, as these two
ultimately set the shape of the luminosity function. The fitting
results are all presented in Table \ref{tab:Schechter_fitsCO}.

 \begin{table}
 \caption{Schechter parameters for the CO J$=$1--0 up to J$=$6--5
   luminosity function from redshift $z=0$ to $z=6$.\label{tab:Schechter_fitsCO}}
 \begin{tabular}{ccccc}
transition & redshift & $\alpha$ & $\log\,L_*$ & $\log\,\phi_*$\\
&&&$\rm{Jy}\,\rm{km}\,\rm{s}^{-1}\,\rm{Mpc}^2$&$\rm{Mpc}^{-3}\,\rm{dex}^{-1}$\\
\hline
 \hline
CO J$=$ 1--0 &0 & $-$1.36 & 6.97 & $-$2.85\\ 
CO J$=$ 1--0 &1 & $-$1.49 & 7.25 & $-$2.73 \\
CO J$=$ 1--0 &2 & $-$1.52 & 7.30 & $-$2.63 \\
CO J$=$ 1--0 &4 & $-$1.71 & 7.26 & $-$2.94 \\
CO J$=$ 1--0 &6 & $-$1.94 & 6.99 & $-$3.46 \\
\hline
CO J$=$ 2--1 &0 & $-$1.35 & 7.54 & $-$2.85 \\
CO J$=$ 2--1 &1 & $-$1.47 & 7.84 & $-$2.72 \\
CO J$=$ 2--1 &2 & $-$1.52 & 7.92 & $-$2.66 \\
CO J$=$ 2--1 &4 & $-$1.75 & 7.89 & $-$3.00 \\
CO J$=$ 2--1 &6 & $-$2.00 & 7.62 & $-$3.56 \\
\hline
CO J$=$ 3--2 &0 & $-$1.29 & 7.83 & $-$2.81 \\
CO J$=$ 3--2 &1 & $-$1.47 & 8.23 & $-$2.79 \\
CO J$=$ 3--2 &2 & $-$1.53 & 8.36 & $-$2.78 \\
CO J$=$ 3--2 &4 & $-$1.76 & 8.26 & $-$3.11 \\
CO J$=$ 3--2 &6 & $-$2.00 & 7.95 & $-$3.60 \\
\hline
CO J$=$ 4--3 &0 & $-$1.29 & 8.16 & $-$2.93 \\
CO J$=$ 4--3 &1 & $-$1.45 & 8.50 & $-$2.84 \\
CO J$=$ 4--3 &2 & $-$1.51 & 8.64 & $-$2.85 \\
CO J$=$ 4--3 &4 & $-$1.80 & 8.70 & $-$3.45 \\
CO J$=$ 4--3 &6 & $-$2.03 & 8.23 & $-$3.78 \\
\hline
CO J$=$ 5--4 &0 & $-$1.20 & 8.37 & $-$2.94 \\
CO J$=$ 5--4 &1 & $-$1.47 & 8.80 & $-$3.03 \\
CO J$=$ 5--4 &2 & $-$1.45 & 8.74 & $-$2.80 \\
CO J$=$ 5--4 &4 & $-$1.76 & 8.73 & $-$3.34 \\
CO J$=$ 5--4 &6 & $-$1.95 & 8.30 & $-$3.67 \\
\hline
CO J$=$ 6--5 &0 & $-$1.15 & 8.38 & $-$2.92 \\
CO J$=$ 6--5 &1 & $-$1.41 & 8.74 & $-$2.92 \\
CO J$=$ 6--5 &2 & $-$1.43 & 8.77 & $-$2.80 \\
CO J$=$ 6--5 &4 & $-$1.73 & 8.84 & $-$3.40 \\
CO J$=$ 6--5 &6 & $-$1.93 & 8.38 & $-$3.72 \\

\hline
\hline
 \end{tabular}
 \end{table}

We plot the evolution of $L_*$ for CO J$=$1--0 out to CO J$=$6--5 in
figure \ref{fig:Lstar_evol}. $L_*$ increases from
$z=6$ to $z=4-3$ for all transitions, after which it gradually
decreases to $z=0$. The evolution in $L_*$ is very minor for CO
J$=$1--0, approximately 0.1-0.2
dex over the entire redshift range probed. CO J$=$2--1 and CO J$=$3--2
have a similar evolution of up to $\sim$0.3 dex. The rate of evolution
increases towards the higher CO J-transitions, where we find a
decrease of 0.5 dex in $L_*$ for CO J$=$6--5 from $z=4$ to $z=0$.

There is a big difference in the absolute value for $L_*$ for the
different CO transitions. Especially CO J$=$1--0 has a characteristic
luminosity 0.5 dex less than CO J$=$2--1 and almost a full dex and even
more for CO J$=$3--2 and higher transitions, respectively.

The right hand panel of Fig. \ref{fig:Lstar_evol} shows the evolution of the powerlaw slope
$\alpha$ for the six different CO J-transitions. We find a general
trend where the slope becomes shallower towards lower redshifts. We
will further discuss these results in Section \ref{sec:discussion}.

The faint end of the CO luminosity functions evolve differently
  for the respective rotational transitions. We predict less evolution
  in the faint end of the CO J$=$1--0  luminosity function than in the
  faint end of for example the CO J$=$4--3 luminosity function. To
  understand this different evolution we plot the \h2-to-CO ratio
   (the
  ratio between molecular hydrogen mass and the velocity
  integrated CO luminosity) as a
  function of CO luminosity for the different rotational transitions in Figure \ref{fig:LCO_alpha}. In general we find that the
  \h2-to-CO ratio decreases with increasing redshift (i.e., the same
  CO luminosity traces a smaller \h2 mass towards higher
  redshifts). Furthermore, we find that the \h2-to-CO ratio decreases
  as a function of CO luminosity. This decline is stronger for high rotational J-transitions. A close look at the
  \h2-to-CO ratios reveals that at CO luminosities of $\sim
  10^6\,\rm{Jy}\,\rm{km}\,\rm{s}^{-1}\,\rm{Mpc}^2$ the ratio between
  \h2 mass and CO J$=$1--0 luminosity evolves with only a factor of
  approximately 2 from redshift $z=6$ to $z=0$, whereas the ratio between \h2 mass
  and CO J$=$6--5 decreases almost 4 times from $z=6$ to $z=0$. % What
  % this effectively means, is that a fixed CO luminosity at $z=6$
  % traces galaxies with an \h2 reservoir a few times less massive than
  % at $z=0$. 
  We will discuss how the
  changing ratio between CO luminosity and \h2 mass shapes the CO
  luminosity functions in Section \ref{sec:discussion}. The predicted turnover
  at low luminosities for luminosity functions of CO rotational
  transitions J$=$3--2 and higher at redshifts $z>2$ is due to
  resolution.

\begin{figure}
 \includegraphics[width = 1.0\hsize]{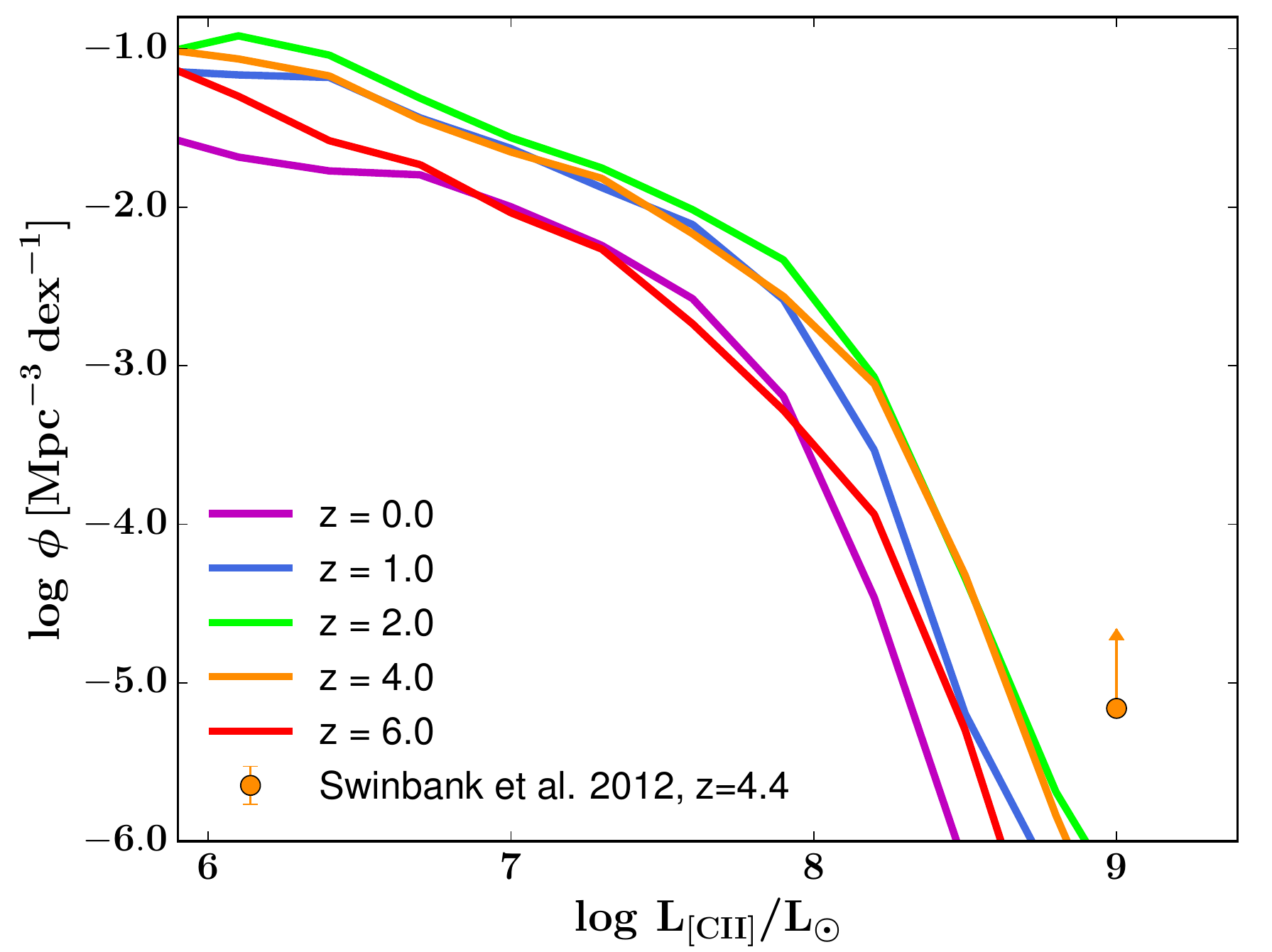}
 \caption{Model predictions of the [CII] luminosity function of
   galaxies from $z=0$ out to $z=6$. A lower limit on the [CII]
   luminosity function at $z=4.4$ is included from \citet{Swinbank2012}.
 \label{fig:CII_lumfunc}}
 \end{figure}

\begin{table}
\caption{Schechter parameters for the [CII] luminosity function from redshift $z=0$ to $z=6$\label{tab:Schechter_fitsCII}}
\begin{tabular}{ccccc}
  transition & redshift & $\alpha$ & $\log\,L_*$ & $\log\,\phi_*$\\
             &&&$\rm{L}_\odot$&$\rm{Mpc}^{-3}\,\rm{dex}^{-1}$\\
  \hline
  \hline
$\rm{[CII]}$&0 & $-$1.25 & 7.47 & $-$2.33\\
$\rm{[CII]}$&1 & $-$1.43 & 7.66 & $-$2.15\\
$\rm{[CII]}$&2 & $-$1.52 & 7.81 & $-$2.20\\
$\rm{[CII]}$&3 & $-$1.41 & 7.80 & $-$2.12\\
$\rm{[CII]}$&4 & $-$1.53 & 7.85 & $-$2.37\\
$\rm{[CII]}$&6 & $-$1.77 & 7.80 & $-$2.95\\
\hline
\hline
 \end{tabular}
 \end{table}

\subsection{[CII] luminosity function}
In Figure \ref{fig:CII_lumfunc} we show the evolution of the [CII]
luminosity function of galaxies from redshift $z=0$ out to $z=6$. We
find a strong evolution in the [CII] luminosity function with
time. We find an increase in the number densities from $z=6$ to
$z=4$. The number densities remain relatively constant from $z=4$ to
$z=2$, and decrease again towards lower redshifts. This behaviour is
again similar to our
predictions for the CO luminosity functions and our predictions for
the \h2 mass functions (PST14), indicative
that our predicted [CII] luminosity function is driven by the same
physical processes.

We compare our predictions with the lower limit set by
\citet{Swinbank2012} based on [CII] observations of two
galaxies at redshift $z\sim4.4$. We find that our predictions for the
[CII] luminosity function at $z=4$ is below the lower limit
found by \citet{Swinbank2012}. This suggests that our model does not
predict enough [CII]-bright galaxies at these redshifts. This is
similar to our predictions for the CO luminosity function at
$z=2.75$, where we barely predicted enough CO bright
objects. We will further discuss the match with the Swinbank et al. lower
limit in Section \ref{sec:discussion}.

\begin{figure*}
 \includegraphics[width = \hsize]{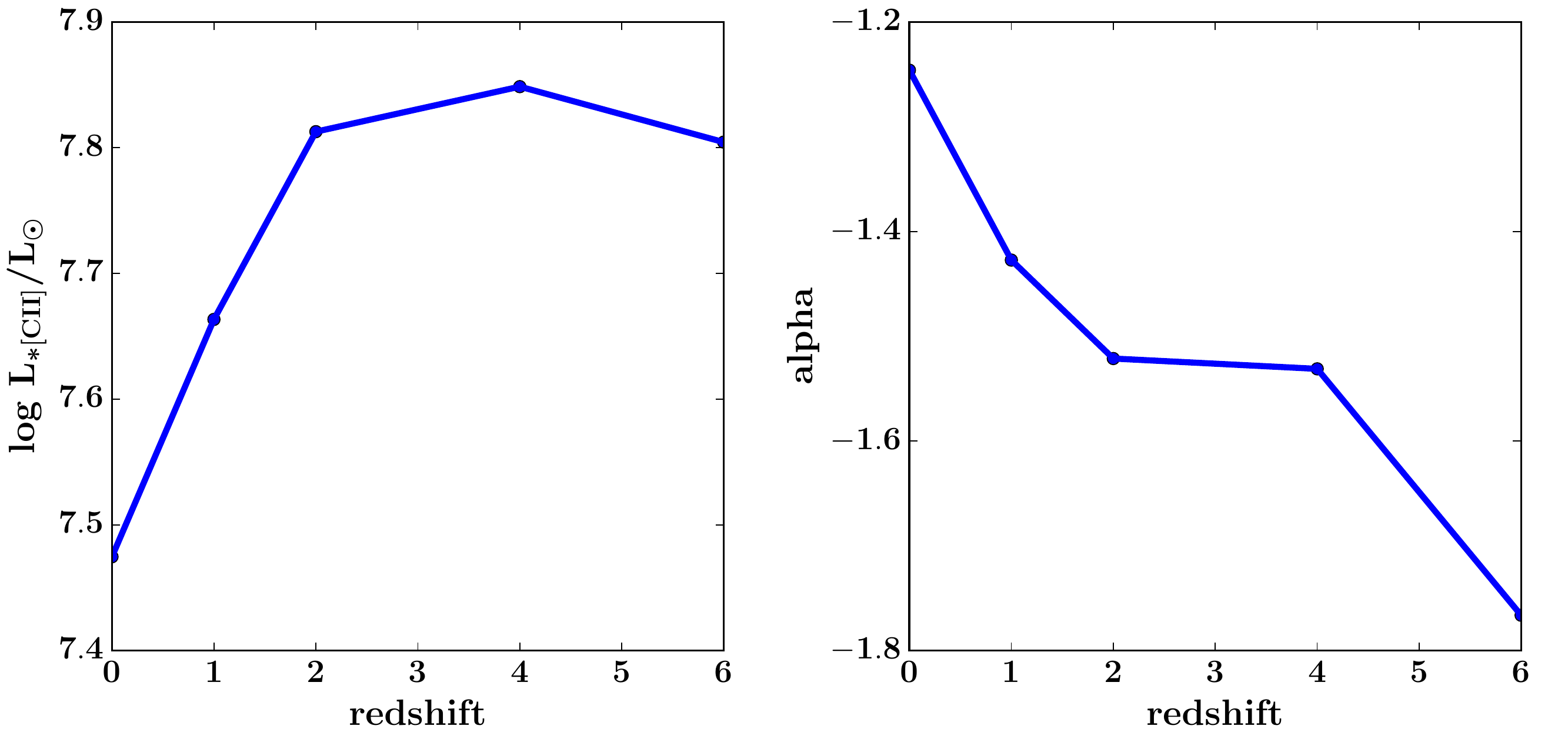}
 \caption{Evolution of the turning point of the Schechter function
   $L_*$ (left) and the slope of the powerlaw component of the Schechter function
   $\alpha$ (right) for the predicted [CII] luminosity function from redshift $z=6$ to $z=0$. \label{fig:Schechter_CII}}
 \end{figure*}

We present the evolution in the parameters 
$L_*{\rm [CII]}$  and $\alpha$ of the Schechter function fit to the [CII] luminosity
function in Figure \ref{fig:Schechter_CII} and in Table \ref{tab:Schechter_fitsCII}. We find that $L_*{\rm
  [CII]}$ is relatively constant from  $z=6$ to $z=2$, and
rapidly decreases at lower redshifts. The rapid drop in $L_*{\rm
  [CII]}$ at redshifts $z<2$ strongly
resembles the evolution of the cosmic SFR density of the Universe,
driven by the strong connection between SFR and [CII] luminosity in
our model. $\alpha$ increases from $z=6$ to $z=4$, then remains relatively
constant out to $z=1.0$, and increases again at later times. 

\begin{table*}
   \begin{center}
   \caption{Survey speed to observe the knee of the CO luminosity
     function for CO J$=$1--0 up to J$=$6--5 over one square degree on
     the sky with a 5 sigma certainty from redshift $z=0$ to
     $z=6$ using ALMA, the JVLA, and the ngVLA. In some cases the line
     is not observable by any of the instrument, in which case the
     required instruments and survey speed are marked with
     n/a. \label{tab:survey_speed}}
 \begin{tabular}{cccccc}
transition & redshift & observed frequency & instrument & rms/pointing & survey
                                                                speed\\
&&GHz&&mJy&hour/arcmin$^2$\\
\hline
 \hline
CO J$=$1--0 & 0 & 115.22 & ALMA band 3& $5\times 10^5$ & 0.03 \\
%CO J$=$1--0 & 0.01 & 114.13 & ALMA band 3& 287.0 & 0.03 \\
%CO J$=$1--0 & 0.3 & 88.67 &ALMA band 3 & 0.37 & 0.2 \\
CO J$=$1--0 & 1 & 67.64 & ALMA band 2& 0.042 & 44.24 \\
CO J$=$1--0 & 2 & 38.42 & ALMA band 1& 0.013 & 20.51 \\
CO J$=$1--0 & 2 & 38.42 & JVLA Ka& 0.013 & 90.00 \\
CO J$=$1--0 & 2 & 38.42 & ngVLA Ka& 0.013 & 18.00 \\
CO J$=$1--0 & 4 & 23.05 & JVLA K& 0.0036 & 239.18 \\
CO J$=$1--0 & 4 & 23.05 & ngVLA K& 0.0036 & 47.83 \\
CO J$=$1--0 & 6 & 16.47 & JVLA Ku & 0.0014 & 341.96\\
CO J$=$1--0 & 6 & 16.47 & ngVLA Ku & 0.0014 & 68.39 \\
\hline
CO J$=$2--1 & 0 & 230.54 &ALMA band 6 & $2\times 10^6$ & 0.12 \\
%CO J$=$2--1 & 0.01 & 228.26 & Alma band 6& 1100.0 & 0.12 \\
CO J$=$2--1 & 1 & 115.26 & ALMA band 3& 0.165 & 4.9 \\
%CO J$=$2--1 & 1.1 & 109.78 & ALMA band 3& 0.139 & 1.93 \\
CO J$=$2--1 & 2 & 76.85 & ALMA band 2& 0.053 & 6.22 \\
CO J$=$2--1 & 4 & 46.11 & JVLA Q& 0.0153 & 132.84 \\
CO J$=$2--1 & 4 & 46.11 & ngVLA Q& 0.0153 & 26.57 \\
CO J$=$2--1 & 6 & 32.93 & ALMA band 1 & 0.0044 & 130.54 \\
CO J$=$2--1 & 6 & 32.93 & JVLA Ka & 0.0044 & 223.5 \\
CO J$=$2--1 & 6 & 32.93 & ngVLA Ka & 0.0044 & 44.70 \\
\hline
CO J$=$3--2 & 0 & 345.8 &ALMA band 7 & $4\times 10^6$ & 0.27 \\
%CO J$=$3--2 & 0.01 & 342.37 &ALMA band 7 & 2100.0 & 0.27 \\
CO J$=$3--2 & 1 & 172.9 & ALMA band 5& 0.201 & 0.73 \\
CO J$=$3--2 & 2 & 115.26 &ALMA band 3 & 0.146 & 6.16 \\
%CO J$=$3--2 & 2.1 & 111.55 &ALMA band 3 & 0.056 & 13.68 \\
CO J$=$3--2 & 4 & 69.16 & ALMA band 2& 0.036 & 28.75 \\
CO J$=$3--2 & 6 & 49.4 &  JVLA Q& 0.0095 & 3345.33 \\
CO J$=$3--2 & 6 & 49.4 &  ngVLA Q& 0.0095 & 669.01 \\
\hline
CO J$=$4--3 & 0 & 461.04 &ALMA band 8 & $7\times 10^6$ & 0.49 \\
%CO J$=$4--3 & 0.01 & 456.48 & ALMA band 8& 4500.0 & 0.48 \\
CO J$=$4--3 & 1 & 230.52 &ALMA band 6 & 0.750 & 0.21 \\
CO J$=$4--3 & 2 & 153.68 & ALMA band 4& 0.278 & 0.74 \\
CO J$=$4--3 & 4 & 92.21 & ALMA band 3& 0.099 & 2.95 \\
%CO J$=$4--3 & 5.7 & 68.81 & ALMA band 2& 0.023 & 79.77 \\
CO J$=$4--3 & 6 & 65.86 & n/a& 0.018 & n/a \\
\hline
CO J$=$5--4 & 0 & 576.27 & n/a& $10^7$ & n/a \\
%CO J$=$5--4 & 0.01 & 570.56 & n/a& 7100.0 & n/a \\
%CO J$=$5--4 & 0.3 & 443.28 & ALMA band 8& 11.29 & 0.45 \\
CO J$=$5--4 & 1 & 288.13 &ALMA band 7 & 1.50 & 0.19 \\
CO J$=$5--4 & 2 & 192.09 &ALMA band 5 & 0.35 & 1.16 \\
CO J$=$5--4 & 4 & 115.25 &ALMA band 3 & 0.106 & 11.66 \\
CO J$=$5--4 & 6 & 82.32 & ALMA band 2& 0.0212 & 36.73 \\
\hline
CO J$=$6--5 & 0 & 691.47 & ALMA band 9& $10^7$ & 1.09 \\
%CO J$=$6--5 & 0.01 & 684.63 &ALMA band 9 & 7300.0 & 1.09 \\
CO J$=$6--5 & 1 & 345.74 & ALMA band 7& 1.304 & 0.27 \\
CO J$=$6--5 & 2 & 230.49 & ALMA band 6& 0.375 & 0.85 \\
CO J$=$6--5 & 4 & 138.29 & ALMA band 4& 0.137 & 2.91 \\
CO J$=$6--5 & 6 & 98.78 & ALMA band 3& 0.0026 & 49.88 \\
\hline
\hline
 \end{tabular}
\end{center}
 \end{table*}

\section{Discussion}
\label{sec:discussion}
\subsection{Observing CO deep fields}
The presented model predictions can be a very valuable asset for
future observing proposals. In Table \ref{tab:survey_speed} we show how much time it requires to detect the knee of the different
CO luminosity functions at our redshifts of
interest over one square arc minute on the sky (the survey speed). Where observable, we
performed the calculations for ALMA (50 twelve
meter antennas), the JVLA, and the ngVLA (assuming dishes of 18
meters). We required a five sigma detection of the knee of the luminosity function (as given in table
\ref{tab:Schechter_fitsCO}) and a spectral resolution of 300
$\rm{km}\,\rm{s}^{-1}$. The reader can use this as a starting point
and easily recalculate the survey speeds for smaller or larger areas or
a different requested sensitivity. The table only takes time on
source into account and one should be aware of additional
overheads. 

We immediately notice that the required observing times vary
significantly. In some cases observing the knee of the luminosity
function with the current instruments only requires a modest
integration time of a few minutes, whereas in other cases it is an
exercise that can easily take up tens of hours. A survey focusing on
the CO J$=$1--0 emission line is much more expensive than surveys
focusing on the higher transitions. This is driven by the strong
difference in characteristic luminosity $L_*$ for CO J$=$1--0 with
respect to the other transitions (see Fig. \ref{fig:Lstar_evol}). 

The CO J$=$3--2 line is
the most favourable transition to observe the global gas content of
galaxies in a deep-field survey during the peak of star-formation of
our Universe. Its survey speed at redshifts $z\leq$ 4 is much shorter
than the survey speeds of the CO J$=$1--0 and CO J$=$2--1 lines. The characteristic density ($\sim
10^{4.5}\,\rm{cm}^{-3}$) of the CO J$=$3--2 line can still be
associated with the bulk molecular gas in a galaxy, which make it more
suitable to observe the molecular reservoir of galaxies than higher
rotational CO lines with higher survey speeds. 

% This is especially helpful towards higher redshifts
% where the CO J$=$1--0 and CO J$=$2--1 emission lines move out of the
% frequency range currently accesible with ALMA. Instrumental
% efficiencies and/or strategies may convince one to point at different lines. 

Though the limited field of view of
ALMA does not make it an ideal survey instrument, its sensitivity
allows one to observe the knee of CO luminosity functions for
high CO rotational transitions at $z=2$ in approximately 10 hours over an area
as big as the Hubble ultra-deep field.

Radio instruments also have the potential to probe
the CO luminosity function of galaxies at redshifts z $>$1, depending on
the exact frequency tunings. The radio regime will become very interesting for
objects towards redshifts of $z>3$, where the CO J$=$3-2 emission line
moves out of the currently available ALMA bands. Our results show that the ngVLA will be much more suitable to carry
out surveys of sub-mm emission lines than the current JVLA. In some
cases the ngVLA is very complementary to ALMA (e.g., to observe CO
J$=$1--0 and CO J$=$2--1 at $z>2$) and in other cases the ngVLA is even more suitable to
observe CO luminosity functions (e.g., the CO J$=$1--0 luminosity
function at $z=2$ and the CO J$=$2--1 luminosity
function at $z=6$). The next generation of radio telescopes
(SKA and its pathfinders ASKAP and MEERKAT) have a very high
sensitivity and large field of view compared with ALMA. If these
instruments are equipped with a high
frequency receiver (targeting frequencies between 1 and 50 GHz) they
will be very efficient carrying out deep fields of low CO J-transitions
at redshifts z $>$ 2. In the near future the ngVLA
is the most obvious telescope to probe low CO rotational
transitions beyond redshifts of $\sim 2$.

We encourage the reader to look for the most favourable frequency
setting when designing a deep-field survey, rather than just focusing
on one CO luminosity function at one particular redshift. With a clever frequency
setting, a limited number of tunings can already probe a number of
different CO luminosity functions at different redshifts
\citep[e.g.,][]{Decarli2014,Walter2014}.

\begin{figure}
\includegraphics[width = \hsize]{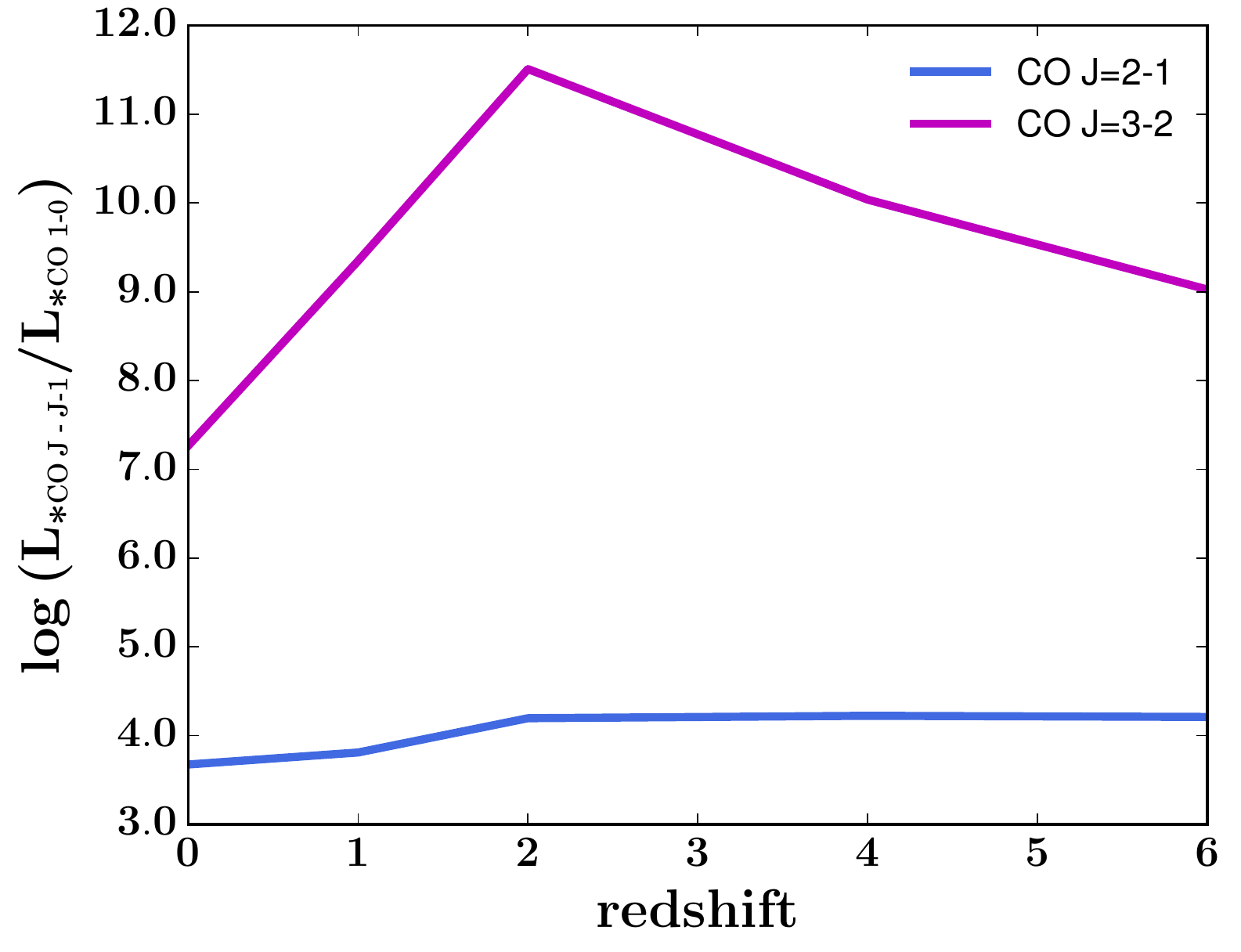}
\caption{The ratio between the characteristic
flux density $L_*$ for CO J$=$2--1 (blue) and CO J$=$3--2 (magenta) and CO J$=$1--0 as a function of redshift. \label{fig:flux_ratio}}
\end{figure}

We want to finish this sub-section with a word of caution. Due to the large difference in rest-frame frequency of the respective
CO J-transitions, current estimates of the CO luminosity function are
based on different CO J-transitions at different redshifts
\citep[uses CO J$=$1--0 at $z=0.0$ and $z=0.3$, CO J$=$2--1 at
$z=1.5$, and CO J$=$3--2 at $z=2.75$]{Walter2014}.  If the goal of a project is to
obtain molecular gas masses, care should be taken to
translate luminosity functions of CO into a CO J$=$1--0 luminosity function. Typically, values of 3.2 and 4.5 are assumed for the flux ratio between CO
J$=$2--1 and CO J$=$3--2, and CO J$=$1--0, respectively \citep[e.g.,][corresponding to brightness temperature luminosity ratios of $L'_{\textrm{CO 2-1}}/L'_{\textrm{CO 1-0}} = 0.8$,
and $L'_{\textrm{CO 3-2}}/L'_{\textrm{CO 1-0}} =
0.5$]{Daddi2014,Dannerbauer2009}.  In Figure
\ref{fig:flux_ratio} we plot the ratio between the characteristic
flux density $L_*$ for the CO J$=$2--1 and CO J$=$3--2 transitions and CO J$=$1--0. At $z=0$
our predictions for the flux ratio are close to the typically adopted
ratios for CO J$=$2--1, and higher for CO J$=$3--2. Our predicted ratio between CO
J$=$2--1 and CO J$=$1--0 remains relatively constant with time. The ratio between CO
J$=$3--2 and CO J$=$1--0 increases towards
higher redshifts and decreases again at redshifts $z>4$. This is driven by changing ISM conditions in
galaxies towards higher redshift (see Figure \ref{fig:ISMprops}), resulting in a larger CO
line ratio
\citep[e.g.,][]{Popping2014radtran,Narayanan2014}. Moreover,
  heating
of the gas by the CMB at high
redshifts can affect the CO line-ratios in
galaxies with low SFRs \citep{Narayanan2014}. Line ratios
can furthermore increase due to the J$=$1--0 line losing contrast with respect
to the CMB background \citep{daCunha2013,Tunnard2016}.

Without
properly accounting for changes in line ratios the number of
galaxies that are bright in CO J$=$1--0 will be overestimated. This
may eventually lead to an incorrect \h2 mass function and an overestimate of the density
of molecular hydrogen in our Universe.

\subsection{Evolution in the shape of the CO luminosity functions}
Our predictions show that the rate of evolution for the characteristic
luminosity $L_*$ is larger for the high CO J-transitions than
for the lower J-transitions (evolution of $\sim 0.1$ dex for CO
J$=$1--0, whereas CO J$=$6--5 evolves with more than $\sim$0.5 dex). This
indicates that not only the predicted amount of total cooling through
CO changes, but also the way this is divided over the different CO
transitions. The notion that less cooling occurs through the higher CO
rotational transitions indicates that the CO bright galaxies also change their
ISM properties, i.e, their ISM consists of a  relatively smaller
component of dense and warm gas. This is in good agreement with
previous predictions made by our models, which showed that as a
function of time the CO Spectral Line Energy Distribution of galaxies
peaks towards lower CO rotational transitions (from redshift $z=2.2$ to
redshifts $z=1.2$, and $z=0.0$; P14). \citet{Daddi2014} demonstrated
that the CO
SLEDs of 2 main-sequence galaxies at $z=1.5$ have an important CO J$=$5--4 component 
not seen in local main-sequence galaxies. This component is also indicative
of clumps of denser and warmer gas in the star-forming ISM of galaxies
at $z=1.5$.

We cannot fail to notice that our models predict the highest
number densities of very bright CO objects at redshifts $z=2-3$. This
coincides with the predicted peak in the cold gas and \h2 cosmic
density (PST14) and the SFR density of our
Universe (SPT15). Within our model the latter is a
natural consequence of the former. A high number density of CO bright objects
is associated with many \h2-rich galaxies. Assuming a molecular-gas
based star-formation relation, this automatically yields a high SFR density.

In Figure \ref{fig:CO_evol} we showed that the  shape of the
CO luminosity function evolves less with redshift  for low rotational
transitions than for higher rotational transitions. We also showed
that at fixed CO luminosity the \h2-to-CO ratio of galaxies
decreases. The evolution in the \h2-to-CO of galaxies is stronger for
the higher than the lower rotational transitions. If a fixed CO luminosity traces a smaller \h2 reservoir at
  high redshift, the volume density that belongs to that CO luminosity reservoir will
  be higher, just because of the slope of the \h2 mass function.
  The evolution in the \h2-to-CO ratio is much
  stronger for the high rotational transitions than for low
  transitions. Therefore, there will be a stronger evolution in
  the volume densities for high rotational CO
  transitions than for low rotational transitions at the faint end of
  the luminosity function. 

These results clearly show that any evolution in the CO luminosity functions is not just
  driven by an evolution in the gas mass, but also by evolution in
the characteristic properties of the ISM that define the shape of the
CO SLEDs as seen in Figure \ref{fig:ISMprops}. 
  Furthermore, the CMB may also influence the shape and evolution of
  the CO luminosity function. Background emission from the CMB can affect the CO luminosity
functions towards higher redshifts, especially the low CO rotational
transitions \citep{Obreschkow2009,daCunha2013,Tunnard2016}. Additional heating of
low-temperature gas by the CMB can slightly increase the excitation
conditions and measured CO intensities \citep{Narayanan2014}.

\subsection{Too few CO-bright galaxies at $z>2$}
We found that our model is barely able to reproduce observational constraints on the  CO J$=$3--2 luminosity
function at $z=2.75$ from the CO blind-survey presented in
\citet[Figure \ref{fig:CO_literature}]{Walter2014}. Walter et al. showed
that a comparison with other semi-analytic models
\citep{Obreschkow2009COSED,Lagos2012} yields similar results. We note that the uncertainties on the
\citet{Walter2014} results are significant. The number of detections
is very limited, and the area on the sky probed very small. Effects of
cosmic variance may have significant influences on the derived CO
number densities. 

\citet{Vallini2016} obtained indirect estimates of the CO luminosity
 function by applying various FIR-to-CO conversions on Herschel
 data. When comparing their empirical estimates of the CO luminosity
 function to model predictions, Vallini et al. also found that theoretical models
 predict too few CO-bright galaxies at $z=2$. Looking at these results a picture emerges where at
$z\geq1.5$ theoretical models predict hardly enough CO-bright
objects.

\begin{figure}
\includegraphics[width = \hsize]{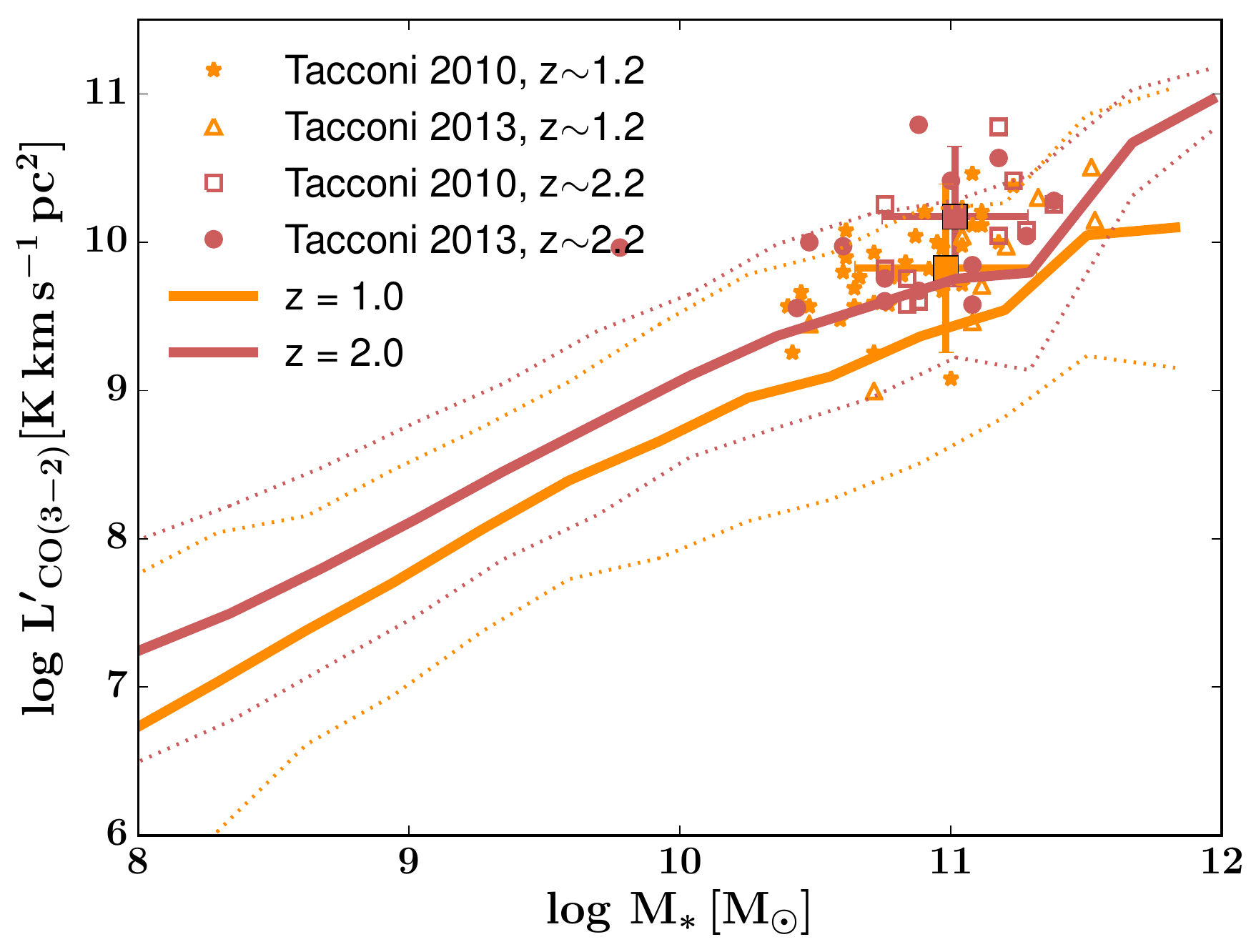}
\caption{CO J$=$3--2 luminosity of galaxies as a function of their
  stellar mass at $z=1$ and $z=2$. Observations are taken from
  \citet{Tacconi2010} and \citet{Tacconi2013}. Large squares with the
  black edges mark the mean CO luminosities at redshift $z=1.2$ and
  $z=2.2$ for the combined samples, respectively.\label{fig:mstar_CO}}
\end{figure}

To further narrow down what could cause the mismatch between our
predictions and the \citet{Walter2014} constraints at $z=1.5$ and
$z=2.75$ we plot the CO J$=$3--2 luminosity of galaxies
as a function of stellar mass at $z=1$ and $z=2$ in Figure \ref{fig:mstar_CO}. We
compare our predictions with observations taken from
\citet{Tacconi2010} and \citet{Tacconi2013} and apply the same
selection criteria to our model galaxies.\footnote{\citet{Tacconi2010}
  and \citet{Tacconi2013} only selected galaxies with SFR $>\,30\,\rm{M}_\odot\,\rm{yr}^{-1}$} We indeed find that our model predictions for the CO
J$=$ 3--2 luminosity of galaxies is approximately 0.3 dex too low at a
given stellar mass,
which could explain the tension between our model predictions and
observational constraints of the CO luminosity function. The
semi-analytic model used in this work matches the observed stellar
mass function at these redshift and at this mass regime quite well
(SPT15), but the faint CO luminosities result in a CO
luminosity function in poor agreement with observations. 

To understand the origin of the mismatch between the predicted and
observed CO luminosity function, we need to take a step back and focus
on the predicted \h2 mass in
galaxies. If we naively assume a constant CO J$=$3--2/CO J$=$1--0 ratio and CO-to-\h2
conversion factor, an underestimation of the CO luminosity of galaxies
by $\sim0.3$ dex will result in an underestimation of the
molecular gas reservoirs of galaxies of 0.3 dex. \citet{Popping2015} extended a sub-halo abundance matching
model with recipes to obtain observationally driven \hi and \h2 masses
of galaxies. They demonstrated that semi-analytic models
that include detailed tracking of atomic and molecular hydrogen
predict $\sim$0.3 dex too little cold gas and \h2 in star-forming galaxies at
$z\sim 2-3$. \citet{Popping2015Candels} inferred the cold gas (\hi + \h2),
\hi, and \h2 gas masses of galaxies taken from the CANDELS
\citep[Cosmic Assembly Near-infrared Deep
Extragalactic Legacy Survey;][]{Grogin2011,Koekemoer2011} survey and also found that
theoretical models predict $\sim 0.3$ dex too little \h2 in galaxies at redshifts
$z=1-3$. A lack of molecular hydrogen translates into
galaxy SFRs that are too low at intermediate redshifts. Indeed,
theoretical models predict $\sim0.3$ dex too little star-formation in
galaxies at $z\sim 2$ with stellar masses $>10^9\,\rm{M}_\odot$ \citep{Somerville2014}. If all other properties of galaxies (stellar mass function,
fraction of star-forming versus quiescent galaxies) are reproduced, too little \h2 will result in an \h2 mass function and CO
luminosity function in poor agreement with
observations. If theoretical models could reproduce the SFRs of
galaxies correctly, presumably the CO line luminosities would be
correct as well.

\citet{Somerville2014} showed that most semi-analytic
and hydrodynamic models fail to reproduce the massive end of the stellar mass function of galaxies
 at $z\sim 2-3$ and predict too few massive galaxies (although models
 that assume that star-formation efficiency increases
 towards high-molecular surface density do much better, SPT15). This naturally
 affects the predicted \h2 mass function and CO luminosity function as
 well. 

The mismatch between predicted CO luminosity function and observations
does not seem to be related to our modelling of the connection between
\h2 and line emission, but to be part of a much
larger set of problems affecting the SFRs and stellar mass growth in
galaxies as well. The results presented in this work are merely a different
representation of this problem, and suggest that too small gas
reservoirs (both molecular as the combination of molecular and atomic
gas) may be at the core of the overall problem. The root of the
problem is in {\it the accretion rate of new gas}, which is modified by
outflows and re-accreting gas. So called 'bathtub models' have
demonstrated the importance of properly accounting for these competing
physical processes \citep{Dave2012,Mitra2015}, Recent work by
\citet{White2014} showed that extending the time for ejected gas to
reaccrete onto galaxies leads to galaxy gas masses in better
agreement with indirect estimates, and improves the match between
predicted and observed SFRs and stellar mass functions \citep[see also][]{Henrigues2014}.

Given the uncertainties discussed above, it is good to ask ourselves the question how far off
the models are from reality and how this affects our
predictions. Galaxy formation models typically reproduce the knee of
the stellar mass function \citep{Somerville2014}. We therefore do
not expect that our predictions will
change much near the knee of the CO luminosity functions. Galaxy
formation models typically predict too few galaxies with stellar masses more
massive than the knee of the stellar mass function. We can thus expect a
higher number of galaxies with CO luminosities brighter than the knee
of the respective CO luminosity functions. We showed that our
model predicts CO J$=$2--1 and CO J$=$3--2 luminosities that are
$\sim$0.3 dex too faint compared to observations (Figure \ref{fig:mstar_CO}), consistent with the low SFRs galaxy formation models predict. If indeed CO luminosities are $\sim$0.3 dex brighter than
suggested by our models, our predicted CO luminosity functions up to
CO J$=$3--2 should be shifted to brighter values by $\sim$0.3 dex. Combined
with the shape of the luminosity functions, these corrections would
shorten the survey speeds mentioned in Tabel \ref{tab:survey_speed} and ease
the observation of the bright end of CO luminosity functions. It is
harder to estimate how our model uncertainties affect the CO
luminosity function of the higher CO rotational transitions, as
excitation physics play a key role. Nevertheless, we expect that a
similar trend holds.

\subsection{The [CII] luminosity function}
% Ionised carbon has often been dubbed the workhorse for extragalactic
% astronomy. Due to its brightness, [CII] is one of the first emission
% lines that can be picked up with sub-mm instruments. This makes it a
% very valuable line to find new objects through blind surveys, or
% assign spectroscopic redshifts \citep[see for a
% review][]{Carilli2013}. [CII] is also considered to be a good tracer of
% star-formation in galaxies \citep{deLooze2011,Herrera2015}, which makes it an extra worthwhile emission
% line to observe.

We found that the [CII] luminosity function remains constant from $z=4$
to $z=2$, after which the number density of the bright [CII]
galaxies decreases. To quantify this we showed that
$L_{*\rm{[CII]}}$, the turning
point between the power-law and exponential component of the Schechter
fit to the [CII] luminosity function, decreases by almost 0.5 dex from $z=2$ to $z=0$. This behaviour is remarkably consistent
with the evolution of the CO luminosity functions presented in this
work and coincides with the predicted peak in the SFR density of our
Universe. It also follows earlier predictions for the \h2 mass function of
galaxies (PST14). This indicates that
these different lines and components are closely correlated. This is
not necessarily surprising. We only account for the
contribution by photo-dissociation regions (PDRs) to the [CII] luminosity of galaxies. These are the same
regions that are responsible for the CO emission and where molecular
hydrogen can form. In reality,
the ionisation of diffuse atomic gas by young stars can also
contribute to the [CII] emission from a galaxy. We will further
discuss this in Section \ref{sec:caveats}. 

We found that our predictions for the [CII] luminosity function at
$z=4$ are somewhat lower than the lower limits derived by
\citet{Swinbank2012}. The lower limits set by \citet{Swinbank2012}
were based on [CII] detections of two galaxies, that were
serendipitously detected as part of a targeted continuum survey on
IR-bright galaxies within a region of 0.25
square degrees. Due to the selection bias and serendipitous nature
  of the detections, the Swinbank et
al. survey may overestimate the number density of [CII]-bright sources
per unit volume. This could explain some of the discrepancy between
the lower limits set in \citet{Swinbank2012} and our work.

\citet{Matsuda2015} combined the data of multiple ALMA Cycle 0 surveys
from the archive to place upper limits on the [CII] luminosity function at
$z\sim4.5$. The upper limits are approximately 3 orders of magnitudes higher than the limits set by Swinbank et al., and do not
constrain our predictions well.

\subsection{Caveats}
\label{sec:caveats}
There are a few physical processes that were not included in this
model which we discuss
here.

\subsubsection{X-ray driven excitation}
Observations with the {\it Herschel Space Observatory} revealed strong excitation of high-J CO lines (CO J$=$9--8 and higher) in nearby
active galaxies \citep{vanderWerf2010,Meijerink2013}. The high
excitation lines can be explained by including the heating from
X-ray radiation on top of the UV radiation. We did not include X-ray heating in our models. We limited the predictions for
our CO luminosity function to CO J$=$6--5, a regime where the
contribution from X-ray heating to the CO luminosity is not thought to
be dominant. The inclusion of X-ray heating could add to the
luminosity of the higher rotational CO transitions such as CO J$=$7--6
and up \citep[see the CO SLEDs in][]{Spaans2008}.

\subsubsection{Mechanical heating and Cosmic rays}
Mechanical heating through shocks increases excitation temperatures
and decreases the optical depth at line centres \citep{Kazandjian2015}. Indeed,
mechanical heating is needed to explain the excitation of
CO in some local luminous infrared galaxies
\citep[e.g.][]{Loenen2008,Meijerink2013,Rosenberg2014a,Rosenberg2014b}. 

A strong cosmic ray field can effectively destroy CO when the cosmic
ray densities are 50 -- 1000 times higher than in our own Milky Way,
affecting the CO luminosity and
CO-to-\h2 conversion rate of galaxies
\citep{Bayet2011,Meijerink2011,Bisbas2015}. This effect may already be
important in Milky Way like giant molecular clouds \citep{Bisbas2015}.

A proper inclusion of the effect of mechanical heating and cosmic rays
(as well as X-ray driven chemistry) would require a much more detailed
chemistry model than currently is applied in this work.

\subsubsection{[CII] emission from ionised regions}
[CII] emission can originate from different phases of the ISM. For
instance, in our own Galaxy 80 per cent of the [CII] comes from
atomic and molecular regions and 20 per cent
from ionised gas \citep{Pineda2014}. For M17SW in the Milky Way the
fraction of [CII] from ionised regions is as high as  33\% \citep{Perez-Beaupuits2015}.

These numbers can change from
galaxy to galaxy and with redshift, depending on the properties of the
ISM in a galaxy. In our work we did not take the contribution from
ionised regions to the [CII] emission of galaxies into
account. \citet{Olsen2015CII} applied a radiative transfer code to
seven modeled main-sequence galaxies at $z=2$. The authors compute the contribution to the
total [CII] emission from PDRs, atomic, and ionised
regions and found that the [CII] emission from ionised regions only accounts for a few percent of the
total [CII] luminosity. 

Observationally the fraction of [CII] emission in extra-galactic
sources arising from ionised
regions is not well
defined. \citet{Decarli2014CII} showed for two Ly-$\alpha$ emitters at
redshift $z=4.7$ that the [CII]-to-[NII] ratio is consistent with the
range of values expected for HII regions. This suggests that most of
the [CII] emission comes from an ionised regime. On the other hand,
Decarli et al. found that the [CII]-to-[NII] ratios in a sub-mm galaxy and quasi-stellar object
at the same redshift are more consistent with a picture where a
substantial fraction of the [CII] emission comes from a neutral
regime. \citet{Gullberg2015} found for 20 dusty star-forming
galaxies that the CO and [CII] emission are consistent with PDR regions. \citet{Cormier2015} showed that [CII] emission from ionised
regions becomes more important towards low-metallicity objects.

These observational results suggest that
while we may be missing the contribution of HII regions in our [CII]
predictions, these are likely not significant at least in the bright end
of the luminosity function.

\section{Summary \& Conclusions}
\label{sec:summary}
In this paper we combined a semi-analytic model of galaxy formation
with a radiative transfer code to make predictions for the evolution of the CO
luminosity function, focusing on the CO J-transitions from J$=$1--0 to
J$=$6--5 and [CII] out to $z=6$. 

\begin{itemize}
\item  Our updated model successfully reproduces the observed scaling
  relations between CO luminosity and FIR luminosity/SFR out to CO
  J$=$6-5 at redshifts $z = 0,\,1$, and $2$, and between [CII]
  luminosity and FIR luminosity. Predicted luminosities
  for CO J$=$ 7--6 up to CO J$=$9--8 are in reasonable agreement with
  observational constraints.

\item We reproduce the observational constraints for the CO luminosity
  function of galaxies at redshifts $z=0$, $z=0.3$, $z=1.5$, and $z=2.75$. 

\item We provide predictions for CO luminosity functions out to
  $z=6$. We find that the number densities of the CO luminosity functions increase from $z=6$ to
  $z=2$, and decrease at lower redshifts. This behaviour is closely
  linked to the history of the SFR density of our Universe. We predict
  that the CO-brightest galaxies can be observed at $z=2$. CO J$=$2--1
  and lower can be picked up by radio instruments, whereas CO J$=$3--2
  and up are ideal to be observed by for instance ALMA and NOEMA.

\item We provide predictions for the [CII] luminosity function of
  galaxies out to $z=6$. Similarly to CO, the [CII] luminosity
  function increases up to $z=2-3$ and decreases at lower redshifts. 

\item Due to its brightness and moderate excitation density, the CO J$=$3--2 emission line is very
  favourable to observe the CO luminosity function and address the
  distribution of molecular gas in our Universe. This line can be
  picked up by ALMA at redshifts z$<$3 and by radio instruments at even
  higher redshifts. Nevertheless, care
  should be taken when converting the CO J$=$3--2 luminosity function
  to a CO J$=$1--0 luminosity function. The ratio between the
  characteristic luminosity describing the turning point between a
  power-law and exponential distribution for these two emission lines
  evolves with redshift.

\item The tension between the CO luminosity function at $z=2.75$ and
  the [CII] luminosity function, and
  the observational constraints may be part of a bigger
  problem. Cosmological simulations have a hard time reproducing the gas
  content and CO emission of galaxies at intermediate redshifts. A
  suitable solution to solve some of the other problems galaxy
  formation models face (mismatch between predicted and observed
  stellar mass functions and sSFR at intermediate redshift) should
  first be able to properly reproduce the gas content of galaxies out
  of which new stars are formed.
\end{itemize}

The results presented in this paper can serve as a theoretical
framework for future deep field efforts with the next generation of
radio and sub-mm instruments.  They provide predictions for such
surveys at the same time. Especially the survey speeds presented in
Table \ref{tab:survey_speed} can be useful for the planning of future
observational efforts. We look forward to future deep field that will be able to confront our predictions and
place more constraints on the physics that drives galaxy formation.

\section*{Acknowledgments}
We thank the referee for a thorough report and very
  constructive comments that have improved the paper. GP thanks Matthieu B\'ethermin, Drew Brisbin, Caitlin Casey, Carl
Ferkinhoff, Alex Karim, Desika
Narayanan, and Fabian Walter for stimulating conversations. RSS thanks the Downsbrough family for their generous support, and acknowledges support from a Simons Investigator award.

\bibliographystyle{mn2e_fix}
\bibliography{references}

\begin{thebibliography}{110}
\expandafter\ifx\csname natexlab\endcsname\relax\def\natexlab#1{#1}\fi

\bibitem[{{Arrigoni} {et~al}\mbox{.}(2010){Arrigoni}, {Trager}, {Somerville},
  \& {Gibson}}]{Arrigoni2010}
{Arrigoni} M., {Trager} S.~C., {Somerville} R.~S., {Gibson} B.~K., 2010,
  \mnras, 402, 173

\bibitem[{{Barnes} {et~al}\mbox{.}(2001){Barnes}, {Staveley-Smith}, {de Blok},
  {Oosterloo}, {Stewart}, {Wright}, {Banks}, {Bhathal}, {Boyce}, {Calabretta},
  {Disney}, {Drinkwater}, {Ekers}, {Freeman}, {Gibson}, {Green}, {Haynes}, {te
  Lintel Hekkert}, {Henning}, {Jerjen}, {Juraszek}, {Kesteven}, {Kilborn},
  {Knezek}, {Koribalski}, {Kraan-Korteweg}, {Malin}, {Marquarding}, {Minchin},
  {Mould}, {Price}, {Putman}, {Ryder}, {Sadler}, {Schr{\"o}der}, {Stootman},
  {Webster}, {Wilson}, \& {Ye}}]{HIPASS}
{Barnes} D.~G. {et~al.}, 2001, \mnras, 322, 486

\bibitem[{{Bayet} {et~al}\mbox{.}(2011){Bayet}, {Williams}, {Hartquist}, \&
  {Viti}}]{Bayet2011}
{Bayet} E., {Williams} D.~A., {Hartquist} T.~W., {Viti} S., 2011, \mnras, 414,
  1583

\bibitem[{{Bigiel} \& {Blitz}(2012)}]{Bigiel2012}
{Bigiel} F., {Blitz} L., 2012, \apj, 756, 183

\bibitem[{{Bigiel} {et~al}\mbox{.}(2008){Bigiel}, {Leroy}, {Walter}, {Brinks},
  {de Blok}, {Madore}, \& {Thornley}}]{Bigiel2008}
{Bigiel} F., {Leroy} A., {Walter} F., {Brinks} E., {de Blok} W.~J.~G., {Madore}
  B., {Thornley} M.~D., 2008, \aj, 136, 2846

\bibitem[{{Bisbas}, {Papadopoulos} \& {Viti}(2015){Bisbas}, {Papadopoulos}, \&
  {Viti}}]{Bisbas2015}
{Bisbas} T.~G., {Papadopoulos} P.~P., {Viti} S., 2015, \apj, 803, 37

\bibitem[{{Blitz} \& {Rosolowsky}(2006)}]{Blitz2006}
{Blitz} L., {Rosolowsky} E., 2006, \apj, 650, 933

\bibitem[{{Blumenthal} {et~al}\mbox{.}(1986){Blumenthal}, {Faber}, {Flores}, \&
  {Primack}}]{Blumenthal1986}
{Blumenthal} G.~R., {Faber} S.~M., {Flores} R., {Primack} J.~R., 1986, \apj,
  301, 27

\bibitem[{{Bolatto}, {Wolfire} \& {Leroy}(2013){Bolatto}, {Wolfire}, \&
  {Leroy}}]{Bolatto2013}
{Bolatto} A.~D., {Wolfire} M., {Leroy} A.~K., 2013, \araa, 51, 207

\bibitem[{{Carilli} {et~al}\mbox{.}(2015){Carilli}, {McKinnon}, {Ott},
  {Beasley}, {Isella}, {Murphy}, {Leroy}, {Casey}, {Moullet}, {Lacy}, {Hodge},
  {Bower}, {Demorest}, {Hull}, {Hughes}, {di Francesco}, {Narayanan}, {Kent},
  {Clark}, \& {Butler}}]{Carilli2015}
{Carilli} C.~L. {et~al.}, 2015, ArXiv e-prints 1510.06438

\bibitem[{{Carilli} \& {Walter}(2013)}]{Carilli2013}
{Carilli} C.~L., {Walter} F., 2013, \araa, 51, 105

\bibitem[{{Casey} {et~al}\mbox{.}(2015){Casey}, {Hodge}, {Lacy}, {Hales},
  {Barger}, {Narayanan}, {Carilli}, {Alatalo}, {da Cunha}, {Emonts}, {Ivison},
  {Kimball}, {Kohno}, {Murphy}, {Riechers}, {Sargent}, \& {Walter}}]{Casey2015}
{Casey} C.~M. {et~al.}, 2015, ArXiv e-prints 1510.06411

\bibitem[{{Cormier} {et~al}\mbox{.}(2015){Cormier}, {Madden}, {Lebouteiller},
  {Abel}, {Hony}, {Galliano}, {R{\'e}my-Ruyer}, {Bigiel}, {Baes}, {Boselli},
  {Chevance}, {Cooray}, {De Looze}, {Doublier}, {Galametz}, {Hughes},
  {Karczewski}, {Lee}, {Lu}, \& {Spinoglio}}]{Cormier2015}
{Cormier} D. {et~al.}, 2015, \aap, 578, A53

\bibitem[{{Crighton} {et~al}\mbox{.}(2015){Crighton}, {Murphy}, {Prochaska},
  {Worseck}, {Rafelski}, {Becker}, {Ellison}, {Fumagalli}, {Lopez}, {Meiksin},
  \& {O'Meara}}]{Crighton2015}
{Crighton} N.~H.~M. {et~al.}, 2015, \mnras, 452, 217

\bibitem[{{da Cunha} {et~al}\mbox{.}(2013){da Cunha}, {Groves}, {Walter},
  {Decarli}, {Weiss}, {Bertoldi}, {Carilli}, {Daddi}, {Elbaz}, {Ivison},
  {Maiolino}, {Riechers}, {Rix}, {Sargent}, \& {Smail}}]{daCunha2013}
{da Cunha} E. {et~al.}, 2013, \apj, 766, 13

\bibitem[{{Daddi} {et~al}\mbox{.}(2010){Daddi}, {Bournaud}, {Walter},
  {Dannerbauer}, {Carilli}, {Dickinson}, {Elbaz}, {Morrison}, {Riechers},
  {Onodera}, {Salmi}, {Krips}, \& {Stern}}]{Daddi2010}
{Daddi} E. {et~al.}, 2010, \apj, 713, 686

\bibitem[{{Daddi} {et~al}\mbox{.}(2015){Daddi}, {Dannerbauer}, {Liu},
  {Aravena}, {Bournaud}, {Walter}, {Riechers}, {Magdis}, {Sargent},
  {B{\'e}thermin}, {Carilli}, {Cibinel}, {Dickinson}, {Elbaz}, {Gao}, {Gobat},
  {Hodge}, \& {Krips}}]{Daddi2014}
---, 2015, \aap, 577, A46

\bibitem[{{Dannerbauer} {et~al}\mbox{.}(2009){Dannerbauer}, {Daddi},
  {Riechers}, {Walter}, {Carilli}, {Dickinson}, {Elbaz}, \&
  {Morrison}}]{Dannerbauer2009}
{Dannerbauer} H., {Daddi} E., {Riechers} D.~A., {Walter} F., {Carilli} C.~L.,
  {Dickinson} M., {Elbaz} D., {Morrison} G.~E., 2009, \apjl, 698, L178

\bibitem[{{Dav{\'e}}, {Finlator} \& {Oppenheimer}(2012){Dav{\'e}}, {Finlator},
  \& {Oppenheimer}}]{Dave2012}
{Dav{\'e}} R., {Finlator} K., {Oppenheimer} B.~D., 2012, \mnras, 421, 98

\bibitem[{{de Looze} {et~al}\mbox{.}(2011){de Looze}, {Baes}, {Bendo},
  {Cortese}, \& {Fritz}}]{deLooze2011}
{de Looze} I., {Baes} M., {Bendo} G.~J., {Cortese} L., {Fritz} J., 2011,
  \mnras, 416, 2712

\bibitem[{{Decarli} {et~al}\mbox{.}(2014{\natexlab{a}}){Decarli}, {Walter},
  {Carilli}, {Bertoldi}, {Cox}, {Ferkinhoff}, {Groves}, {Maiolino}, {Neri},
  {Riechers}, \& {Weiss}}]{Decarli2014CII}
{Decarli} R. {et~al.}, 2014{\natexlab{a}}, \apjl, 782, L17

\bibitem[{{Decarli} {et~al}\mbox{.}(2014{\natexlab{b}}){Decarli}, {Walter},
  {Carilli}, {Riechers}, {Cox}, {Neri}, {Aravena}, {Bell}, {Bertoldi},
  {Colombo}, {Da Cunha}, {Daddi}, {Dickinson}, {Downes}, {Ellis}, {Lentati},
  {Maiolino}, {Menten}, {Rix}, {Sargent}, {Stark}, {Weiner}, \&
  {Weiss}}]{Decarli2014}
---, 2014{\natexlab{b}}, \apj, 782, 78

\bibitem[{{Eales} {et~al}\mbox{.}(2010){Eales}, {Dunne}, {Clements}, {Cooray},
  {de Zotti}, {Dye}, {Ivison}, {Jarvis}, {Lagache}, {Maddox}, {Negrello},
  {Serjeant}, {Thompson}, {van Kampen}, {Amblard}, {Andreani}, {Baes},
  {Beelen}, {Bendo}, {Benford}, {Bertoldi}, {Bock}, {Bonfield}, {Boselli},
  {Bridge}, {Buat}, {Burgarella}, {Carlberg}, {Cava}, {Chanial}, {Charlot},
  {Christopher}, {Coles}, {Cortese}, {Dariush}, {da Cunha}, {Dalton}, {Danese},
  {Dannerbauer}, {Driver}, {Dunlop}, {Fan}, {Farrah}, {Frayer}, {Frenk},
  {Geach}, {Gardner}, {Gomez}, {Gonz{\'a}lez-Nuevo}, {Gonz{\'a}lez-Solares},
  {Griffin}, {Hardcastle}, {Hatziminaoglou}, {Herranz}, {Hughes}, {Ibar},
  {Jeong}, {Lacey}, {Lapi}, {Lawrence}, {Lee}, {Leeuw}, {Liske},
  {L{\'o}pez-Caniego}, {M{\"u}ller}, {Nandra}, {Panuzzo}, {Papageorgiou},
  {Patanchon}, {Peacock}, {Pearson}, {Phillipps}, {Pohlen}, {Popescu},
  {Rawlings}, {Rigby}, {Rigopoulou}, {Robotham}, {Rodighiero}, {Sansom},
  {Schulz}, {Scott}, {Smith}, {Sibthorpe}, {Smail}, {Stevens}, {Sutherland},
  {Takeuchi}, {Tedds}, {Temi}, {Tuffs}, {Trichas}, {Vaccari}, {Valtchanov},
  {van der Werf}, {Verma}, {Vieria}, {Vlahakis}, \& {White}}]{Hatlas2010}
{Eales} S. {et~al.}, 2010, \pasp, 122, 499

\bibitem[{{Federrath}, {Klessen} \& {Schmidt}(2008){Federrath}, {Klessen}, \&
  {Schmidt}}]{Federrath2008}
{Federrath} C., {Klessen} R.~S., {Schmidt} W., 2008, \apjl, 688, L79

\bibitem[{{Feldmann}, {Gnedin} \& {Kravtsov}(2012){Feldmann}, {Gnedin}, \&
  {Kravtsov}}]{Feldmann2012}
{Feldmann} R., {Gnedin} N.~Y., {Kravtsov} A.~V., 2012, \apj, 747, 124

\bibitem[{{Flores} {et~al}\mbox{.}(1993){Flores}, {Primack}, {Blumenthal}, \&
  {Faber}}]{Flores1993}
{Flores} R., {Primack} J.~R., {Blumenthal} G.~R., {Faber} S.~M., 1993, \apj,
  412, 443

\bibitem[{{Genzel} {et~al}\mbox{.}(2010){Genzel}, {Tacconi}, {Gracia-Carpio},
  {Sternberg}, {Cooper}, {Shapiro}, {Bolatto}, {Bouch{\'e}}, {Bournaud},
  {Burkert}, {Combes}, {Comerford}, {Cox}, {Davis}, {Schreiber},
  {Garcia-Burillo}, {Lutz}, {Naab}, {Neri}, {Omont}, {Shapley}, \&
  {Weiner}}]{Genzel2010}
{Genzel} R. {et~al.}, 2010, \mnras, 407, 2091

\bibitem[{{Giovanelli} {et~al}\mbox{.}(2005){Giovanelli}, {Haynes}, {Kent},
  {Perillat}, {Saintonge}, {Brosch}, {Catinella}, {Hoffman}, {Stierwalt},
  {Spekkens}, {Lerner}, {Masters}, {Momjian}, {Rosenberg}, {Springob},
  {Boselli}, {Charmandaris}, {Darling}, {Davies}, {Garcia Lambas}, {Gavazzi},
  {Giovanardi}, {Hardy}, {Hunt}, {Iovino}, {Karachentsev}, {Karachentseva},
  {Koopmann}, {Marinoni}, {Minchin}, {Muller}, {Putman}, {Pantoja}, {Salzer},
  {Scodeggio}, {Skillman}, {Solanes}, {Valotto}, {van Driel}, \& {van
  Zee}}]{ALFALFA}
{Giovanelli} R. {et~al.}, 2005, \aj, 130, 2598

\bibitem[{{Gnedin} \& {Kravtsov}(2011)}]{Gnedin2011}
{Gnedin} N.~Y., {Kravtsov} A.~V., 2011, \apj, 728, 88

\bibitem[{{Greve} {et~al}\mbox{.}(2014){Greve}, {Leonidaki}, {Xilouris},
  {Wei{\ss}}, {Zhang}, {van der Werf}, {Aalto}, {Armus}, {D{\'{\i}}az-Santos},
  {Evans}, {Fischer}, {Gao}, {Gonz{\'a}lez-Alfonso}, {Harris}, {Henkel},
  {Meijerink}, {Naylor}, {Smith}, {Spaans}, {Stacey}, {Veilleux}, \&
  {Walter}}]{Greve2014}
{Greve} T.~R. {et~al.}, 2014, \apj, 794, 142

\bibitem[{{Grogin} {et~al}\mbox{.}(2011){Grogin}, {Kocevski}, {Faber},
  {Ferguson}, {Koekemoer}, {Riess}, {Acquaviva}, {Alexander}, {Almaini},
  {Ashby}, {Barden}, {Bell}, {Bournaud}, {Brown}, {Caputi}, {Casertano},
  {Cassata}, {Castellano}, {Challis}, {Chary}, {Cheung}, {Cirasuolo},
  {Conselice}, {Roshan Cooray}, {Croton}, {Daddi}, {Dahlen}, {Dav{\'e}}, {de
  Mello}, {Dekel}, {Dickinson}, {Dolch}, {Donley}, {Dunlop}, {Dutton}, {Elbaz},
  {Fazio}, {Filippenko}, {Finkelstein}, {Fontana}, {Gardner}, {Garnavich},
  {Gawiser}, {Giavalisco}, {Grazian}, {Guo}, {Hathi}, {H{\"a}ussler},
  {Hopkins}, {Huang}, {Huang}, {Jha}, {Kartaltepe}, {Kirshner}, {Koo}, {Lai},
  {Lee}, {Li}, {Lotz}, {Lucas}, {Madau}, {McCarthy}, {McGrath}, {McIntosh},
  {McLure}, {Mobasher}, {Moustakas}, {Mozena}, {Nandra}, {Newman}, {Niemi},
  {Noeske}, {Papovich}, {Pentericci}, {Pope}, {Primack}, {Rajan},
  {Ravindranath}, {Reddy}, {Renzini}, {Rix}, {Robaina}, {Rodney}, {Rosario},
  {Rosati}, {Salimbeni}, {Scarlata}, {Siana}, {Simard}, {Smidt}, {Somerville},
  {Spinrad}, {Straughn}, {Strolger}, {Telford}, {Teplitz}, {Trump}, {van der
  Wel}, {Villforth}, {Wechsler}, {Weiner}, {Wiklind}, {Wild}, {Wilson},
  {Wuyts}, {Yan}, \& {Yun}}]{Grogin2011}
{Grogin} N.~A. {et~al.}, 2011, \apjs, 197, 35

\bibitem[{{Gullberg} {et~al}\mbox{.}(2015){Gullberg}, {De Breuck}, {Vieira},
  {Wei{\ss}}, {Aguirre}, {Aravena}, {B{\'e}thermin}, {Bradford}, {Bothwell},
  {Carlstrom}, {Chapman}, {Fassnacht}, {Gonzalez}, {Greve}, {Hezaveh},
  {Holzapfel}, {Husband}, {Ma}, {Malkan}, {Marrone}, {Menten}, {Murphy},
  {Reichardt}, {Spilker}, {Stark}, {Strandet}, \& {Welikala}}]{Gullberg2015}
{Gullberg} B. {et~al.}, 2015, \mnras, 449, 2883

\bibitem[{{Henriques} {et~al}\mbox{.}(2015){Henriques}, {White}, {Thomas},
  {Angulo}, {Guo}, {Lemson}, {Springel}, \& {Overzier}}]{Henrigues2014}
{Henriques} B.~M.~B., {White} S.~D.~M., {Thomas} P.~A., {Angulo} R., {Guo} Q.,
  {Lemson} G., {Springel} V., {Overzier} R., 2015, \mnras, 451, 2663

\bibitem[{{Herrera-Camus} {et~al}\mbox{.}(2015){Herrera-Camus}, {Bolatto},
  {Wolfire}, {Smith}, {Croxall}, {Kennicutt}, {Calzetti}, {Helou}, {Walter},
  {Leroy}, {Draine}, {Brandl}, {Armus}, {Sandstrom}, {Dale}, {Aniano}, {Meidt},
  {Boquien}, {Hunt}, {Galametz}, {Tabatabaei}, {Murphy}, {Appleton}, {Roussel},
  {Engelbracht}, \& {Beirao}}]{Herrera2015}
{Herrera-Camus} R. {et~al.}, 2015, \apj, 800, 1

\bibitem[{{Hodge} {et~al}\mbox{.}(2015){Hodge}, {Riechers}, {Decarli},
  {Walter}, {Carilli}, {Daddi}, \& {Dannerbauer}}]{Hodge2014}
{Hodge} J.~A., {Riechers} D., {Decarli} R., {Walter} F., {Carilli} C.~L.,
  {Daddi} E., {Dannerbauer} H., 2015, \apjl, 798, L18

\bibitem[{{Hollenbach}, {Takahashi} \& {Tielens}(1991){Hollenbach},
  {Takahashi}, \& {Tielens}}]{Hollenbach1991}
{Hollenbach} D.~J., {Takahashi} T., {Tielens} A.~G.~G.~M., 1991, \apj, 377, 192

\bibitem[{{Hopkins}(2004)}]{Hopkins2004}
{Hopkins} A.~M., 2004, \apj, 615, 209

\bibitem[{{Hopkins} \& {Beacom}(2006)}]{Hopkins2006}
{Hopkins} A.~M., {Beacom} J.~F., 2006, \apj, 651, 142

\bibitem[{{Juneau} {et~al}\mbox{.}(2009){Juneau}, {Narayanan}, {Moustakas},
  {Shirley}, {Bussmann}, {Kennicutt}, \& {Vanden Bout}}]{Juneau2009}
{Juneau} S., {Narayanan} D.~T., {Moustakas} J., {Shirley} Y.~L., {Bussmann}
  R.~S., {Kennicutt}, Jr. R.~C., {Vanden Bout} P.~A., 2009, \apj, 707, 1217

\bibitem[{{Kamenetzky} {et~al}\mbox{.}(2015){Kamenetzky}, {Rangwala}, {Glenn},
  {Maloney}, \& {Conley}}]{Kamenetzky2016}
{Kamenetzky} J., {Rangwala} N., {Glenn} J., {Maloney} P.~R., {Conley} A., 2015,
  ArXiv e-prints 1508.05102

\bibitem[{{Kazandjian} {et~al}\mbox{.}(2015){Kazandjian}, {Meijerink},
  {Pelupessy}, {Israel}, \& {Spaans}}]{Kazandjian2015}
{Kazandjian} M.~V., {Meijerink} R., {Pelupessy} I., {Israel} F.~P., {Spaans}
  M., 2015, \aap, 574, A127

\bibitem[{{Keres}, {Yun} \& {Young}(2003){Keres}, {Yun}, \&
  {Young}}]{Keres2003}
{Keres} D., {Yun} M.~S., {Young} J.~S., 2003, \apj, 582, 659

\bibitem[{{Koekemoer} {et~al}\mbox{.}(2011){Koekemoer}, {Faber}, {Ferguson},
  {Grogin}, {Kocevski}, {Koo}, {Lai}, {Lotz}, {Lucas}, {McGrath}, {Ogaz},
  {Rajan}, {Riess}, {Rodney}, {Strolger}, {Casertano}, {Castellano}, {Dahlen},
  {Dickinson}, {Dolch}, {Fontana}, {Giavalisco}, {Grazian}, {Guo}, {Hathi},
  {Huang}, {van der Wel}, {Yan}, {Acquaviva}, {Alexander}, {Almaini}, {Ashby},
  {Barden}, {Bell}, {Bournaud}, {Brown}, {Caputi}, {Cassata}, {Challis},
  {Chary}, {Cheung}, {Cirasuolo}, {Conselice}, {Roshan Cooray}, {Croton},
  {Daddi}, {Dav{\'e}}, {de Mello}, {de Ravel}, {Dekel}, {Donley}, {Dunlop},
  {Dutton}, {Elbaz}, {Fazio}, {Filippenko}, {Finkelstein}, {Frazer}, {Gardner},
  {Garnavich}, {Gawiser}, {Gruetzbauch}, {Hartley}, {H{\"a}ussler},
  {Herrington}, {Hopkins}, {Huang}, {Jha}, {Johnson}, {Kartaltepe},
  {Khostovan}, {Kirshner}, {Lani}, {Lee}, {Li}, {Madau}, {McCarthy},
  {McIntosh}, {McLure}, {McPartland}, {Mobasher}, {Moreira}, {Mortlock},
  {Moustakas}, {Mozena}, {Nandra}, {Newman}, {Nielsen}, {Niemi}, {Noeske},
  {Papovich}, {Pentericci}, {Pope}, {Primack}, {Ravindranath}, {Reddy},
  {Renzini}, {Rix}, {Robaina}, {Rosario}, {Rosati}, {Salimbeni}, {Scarlata},
  {Siana}, {Simard}, {Smidt}, {Snyder}, {Somerville}, {Spinrad}, {Straughn},
  {Telford}, {Teplitz}, {Trump}, {Vargas}, {Villforth}, {Wagner}, {Wandro},
  {Wechsler}, {Weiner}, {Wiklind}, {Wild}, {Wilson}, {Wuyts}, \&
  {Yun}}]{Koekemoer2011}
{Koekemoer} A.~M. {et~al.}, 2011, \apjs, 197, 36

\bibitem[{{Komatsu} {et~al}\mbox{.}(2009){Komatsu}, {Dunkley}, {Nolta},
  {Bennett}, {Gold}, {Hinshaw}, {Jarosik}, {Larson}, {Limon}, {Page},
  {Spergel}, {Halpern}, {Hill}, {Kogut}, {Meyer}, {Tucker}, {Weiland},
  {Wollack}, \& {Wright}}]{Komatsu2009}
{Komatsu} E. {et~al.}, 2009, \apjs, 180, 330

\bibitem[{{Kravtsov}(1999)}]{Kravtsov99}
{Kravtsov} A.~V., 1999, PhD thesis, NEW MEXICO STATE UNIVERSITY

\bibitem[{{Krumholz}(2014)}]{Krumholz2013}
{Krumholz} M.~R., 2014, \mnras, 437, 1662

\bibitem[{{Lagos} {et~al}\mbox{.}(2012){Lagos}, {Bayet}, {Baugh}, {Lacey},
  {Bell}, {Fanidakis}, \& {Geach}}]{Lagos2012}
{Lagos} C.~d.~P., {Bayet} E., {Baugh} C.~M., {Lacey} C.~G., {Bell} T.~A.,
  {Fanidakis} N., {Geach} J.~E., 2012, \mnras, 426, 2142

\bibitem[{{Lang} {et~al}\mbox{.}(2014){Lang}, {Wuyts}, {Somerville},
  {F{\"o}rster Schreiber}, {Genzel}, {Bell}, {Brammer}, {Dekel}, {Faber},
  {Ferguson}, {Grogin}, {Kocevski}, {Koekemoer}, {Lutz}, {McGrath}, {Momcheva},
  {Nelson}, {Primack}, {Rosario}, {Skelton}, {Tacconi}, {van Dokkum}, \&
  {Whitaker}}]{Lang2014}
{Lang} P. {et~al.}, 2014, \apj, 788, 11

\bibitem[{{Lemaster} \& {Stone}(2008)}]{Lemaster2008}
{Lemaster} M.~N., {Stone} J.~M., 2008, \apjl, 682, L97

\bibitem[{{Leroy} {et~al}\mbox{.}(2009){Leroy}, {Walter}, {Bigiel}, {Usero},
  {Weiss}, {Brinks}, {de Blok}, {Kennicutt}, {Schuster}, {Kramer},
  {Wiesemeyer}, \& {Roussel}}]{Leroy2009CO}
{Leroy} A.~K. {et~al.}, 2009, \aj, 137, 4670

\bibitem[{{Leroy} {et~al}\mbox{.}(2008){Leroy}, {Walter}, {Brinks}, {Bigiel},
  {de Blok}, {Madore}, \& {Thornley}}]{Leroy2008}
{Leroy} A.~K., {Walter} F., {Brinks} E., {Bigiel} F., {de Blok} W.~J.~G.,
  {Madore} B., {Thornley} M.~D., 2008, \aj, 136, 2782

\bibitem[{{Lisenfeld} {et~al}\mbox{.}(2011){Lisenfeld}, {Espada},
  {Verdes-Montenegro}, {Kuno}, {Leon}, {Sabater}, {Sato}, {Sulentic}, {Verley},
  \& {Yun}}]{Lisenfeld2011}
{Lisenfeld} U. {et~al.}, 2011, \aap, 534, A102

\bibitem[{{Liu} {et~al}\mbox{.}(2015){Liu}, {Gao}, {Isaak}, {Daddi}, {Yang},
  {Lu}, \& {van der Werf}}]{Liu2015}
{Liu} D., {Gao} Y., {Isaak} K., {Daddi} E., {Yang} C., {Lu} N., {van der Werf}
  P., 2015, \apjl, 810, L14

\bibitem[{{Loenen} {et~al}\mbox{.}(2008){Loenen}, {Spaans}, {Baan}, \&
  {Meijerink}}]{Loenen2008}
{Loenen} A.~F., {Spaans} M., {Baan} W.~A., {Meijerink} R., 2008, \aap, 488, L5

\bibitem[{{Madau} \& {Dickinson}(2014)}]{Madau2014}
{Madau} P., {Dickinson} M., 2014, \araa, 52, 415

\bibitem[{{Madau} {et~al}\mbox{.}(1996){Madau}, {Ferguson}, {Dickinson},
  {Giavalisco}, {Steidel}, \& {Fruchter}}]{Madau1996}
{Madau} P., {Ferguson} H.~C., {Dickinson} M.~E., {Giavalisco} M., {Steidel}
  C.~C., {Fruchter} A., 1996, \mnras, 283, 1388

\bibitem[{{Matsuda} {et~al}\mbox{.}(2015){Matsuda}, {Nagao}, {Iono},
  {Hatsukade}, {Kohno}, {Tamura}, {Yamaguchi}, \& {Shimizu}}]{Matsuda2015}
{Matsuda} Y., {Nagao} T., {Iono} D., {Hatsukade} B., {Kohno} K., {Tamura} Y.,
  {Yamaguchi} Y., {Shimizu} I., 2015, \mnras, 451, 1141

\bibitem[{{Meijerink} {et~al}\mbox{.}(2013){Meijerink}, {Kristensen},
  {Wei{\ss}}, {van der Werf}, {Walter}, {Spaans}, {Loenen}, {Fischer},
  {Israel}, {Isaak}, {Papadopoulos}, {Aalto}, {Armus}, {Charmandaris},
  {Dasyra}, {Diaz-Santos}, {Evans}, {Gao}, {Gonz{\'a}lez-Alfonso},
  {G{\"u}sten}, {Henkel}, {Kramer}, {Lord}, {Mart{\'{\i}}n-Pintado}, {Naylor},
  {Sanders}, {Smith}, {Spinoglio}, {Stacey}, {Veilleux}, \&
  {Wiedner}}]{Meijerink2013}
{Meijerink} R. {et~al.}, 2013, \apjl, 762, L16

\bibitem[{{Meijerink} \& {Spaans}(2005)}]{Meijerink2005}
{Meijerink} R., {Spaans} M., 2005, \aap, 436, 397

\bibitem[{{Meijerink} {et~al}\mbox{.}(2011){Meijerink}, {Spaans}, {Loenen}, \&
  {van der Werf}}]{Meijerink2011}
{Meijerink} R., {Spaans} M., {Loenen} A.~F., {van der Werf} P.~P., 2011, \aap,
  525, A119

\bibitem[{{Mitra}, {Dav{\'e}} \& {Finlator}(2015){Mitra}, {Dav{\'e}}, \&
  {Finlator}}]{Mitra2015}
{Mitra} S., {Dav{\'e}} R., {Finlator} K., 2015, \mnras, 452, 1184

\bibitem[{{Mo}, {Mao} \& {White}(1998){Mo}, {Mao}, \& {White}}]{Mo1998}
{Mo} H.~J., {Mao} S., {White} S.~D.~M., 1998, \mnras, 295, 319

\bibitem[{{Murray} \& {Rahman}(2010)}]{Murray2010}
{Murray} N., {Rahman} M., 2010, \apj, 709, 424

\bibitem[{{Narayanan} {et~al}\mbox{.}(2008){Narayanan}, {Cox}, {Kelly},
  {Dav{\'e}}, {Hernquist}, {Di Matteo}, {Hopkins}, {Kulesa}, {Robertson}, \&
  {Walker}}]{Narayanan2008}
{Narayanan} D. {et~al.}, 2008, \apjs, 176, 331

\bibitem[{{Narayanan} {et~al}\mbox{.}(2005){Narayanan}, {Groppi}, {Kulesa}, \&
  {Walker}}]{Narayanan2005}
{Narayanan} D., {Groppi} C.~E., {Kulesa} C.~A., {Walker} C.~K., 2005, \apj,
  630, 269

\bibitem[{{Narayanan} \& {Krumholz}(2014)}]{Narayanan2014}
{Narayanan} D., {Krumholz} M.~R., 2014, \mnras, 442, 1411

\bibitem[{{Obreschkow} {et~al}\mbox{.}(2009){Obreschkow}, {Heywood},
  {Kl{\"o}ckner}, \& {Rawlings}}]{Obreschkow2009COSED}
{Obreschkow} D., {Heywood} I., {Kl{\"o}ckner} H.-R., {Rawlings} S., 2009, \apj,
  702, 1321

\bibitem[{{Obreschkow} \& {Rawlings}(2009)}]{Obreschkow2009}
{Obreschkow} D., {Rawlings} S., 2009, \mnras, 394, 1857

\bibitem[{{Oliver} {et~al}\mbox{.}(2012){Oliver}, {Bock}, {Altieri}, {Amblard},
  {Arumugam}, {Aussel}, {Babbedge}, {Beelen}, {B{\'e}thermin}, {Blain},
  {Boselli}, {Bridge}, {Brisbin}, {Buat}, {Burgarella},
  {Castro-Rodr{\'{\i}}guez}, {Cava}, {Chanial}, {Cirasuolo}, {Clements},
  {Conley}, {Conversi}, {Cooray}, {Dowell}, {Dubois}, {Dwek}, {Dye}, {Eales},
  {Elbaz}, {Farrah}, {Feltre}, {Ferrero}, {Fiolet}, {Fox}, {Franceschini},
  {Gear}, {Giovannoli}, {Glenn}, {Gong}, {Gonz{\'a}lez Solares}, {Griffin},
  {Halpern}, {Harwit}, {Hatziminaoglou}, {Heinis}, {Hurley}, {Hwang}, {Hyde},
  {Ibar}, {Ilbert}, {Isaak}, {Ivison}, {Lagache}, {Le Floc'h}, {Levenson},
  {Faro}, {Lu}, {Madden}, {Maffei}, {Magdis}, {Mainetti}, {Marchetti},
  {Marsden}, {Marshall}, {Mortier}, {Nguyen}, {O'Halloran}, {Omont}, {Page},
  {Panuzzo}, {Papageorgiou}, {Patel}, {Pearson}, {P{\'e}rez-Fournon}, {Pohlen},
  {Rawlings}, {Raymond}, {Rigopoulou}, {Riguccini}, {Rizzo}, {Rodighiero},
  {Roseboom}, {Rowan-Robinson}, {S{\'a}nchez Portal}, {Schulz}, {Scott},
  {Seymour}, {Shupe}, {Smith}, {Stevens}, {Symeonidis}, {Trichas}, {Tugwell},
  {Vaccari}, {Valtchanov}, {Vieira}, {Viero}, {Vigroux}, {Wang}, {Ward},
  {Wardlow}, {Wright}, {Xu}, \& {Zemcov}}]{Hermes2012}
{Oliver} S.~J. {et~al.}, 2012, \mnras, 424, 1614

\bibitem[{{Olsen} {et~al}\mbox{.}(2015{\natexlab{a}}){Olsen}, {Greve},
  {Brinch}, {Sommer-Larsen}, {Rasmussen}, {Toft}, \& {Zirm}}]{Olsen2015CO}
{Olsen} K.~P., {Greve} T.~R., {Brinch} C., {Sommer-Larsen} J., {Rasmussen} J.,
  {Toft} S., {Zirm} A., 2015{\natexlab{a}}, ArXiv e-prints 1507.00012

\bibitem[{{Olsen} {et~al}\mbox{.}(2015{\natexlab{b}}){Olsen}, {Greve},
  {Narayanan}, {Thompson}, {Toft}, \& {Brinch}}]{Olsen2015CII}
{Olsen} K.~P., {Greve} T.~R., {Narayanan} D., {Thompson} R., {Toft} S.,
  {Brinch} C., 2015{\natexlab{b}}, \apj, 814, 76

\bibitem[{{Ostriker}, {Stone} \& {Gammie}(2001){Ostriker}, {Stone}, \&
  {Gammie}}]{Ostriker2001}
{Ostriker} E.~C., {Stone} J.~M., {Gammie} C.~F., 2001, \apj, 546, 980

\bibitem[{{Papadopoulos} {et~al}\mbox{.}(2012){Papadopoulos}, {van der Werf},
  {Xilouris}, {Isaak}, {Gao}, \& {M{\"u}hle}}]{Papadopoulos2012}
{Papadopoulos} P.~P., {van der Werf} P.~P., {Xilouris} E.~M., {Isaak} K.~G.,
  {Gao} Y., {M{\"u}hle} S., 2012, \mnras, 426, 2601

\bibitem[{{P{\'e}rez-Beaupuits} {et~al}\mbox{.}(2015){P{\'e}rez-Beaupuits},
  {Stutzki}, {Ossenkopf}, {Spaans}, {G{\"u}sten}, \&
  {Wiesemeyer}}]{Perez-Beaupuits2015}
{P{\'e}rez-Beaupuits} J.~P., {Stutzki} J., {Ossenkopf} V., {Spaans} M.,
  {G{\"u}sten} R., {Wiesemeyer} H., 2015, \aap, 575, A9

\bibitem[{{P{\'e}rez-Beaupuits}, {Wada} \& {Spaans}(2011){P{\'e}rez-Beaupuits},
  {Wada}, \& {Spaans}}]{JP2011}
{P{\'e}rez-Beaupuits} J.~P., {Wada} K., {Spaans} M., 2011, \apj, 730, 48

\bibitem[{{Pilbratt} {et~al}\mbox{.}(2010){Pilbratt}, {Riedinger}, {Passvogel},
  {Crone}, {Doyle}, {Gageur}, {Heras}, {Jewell}, {Metcalfe}, {Ott}, \&
  {Schmidt}}]{Pilbratt2010}
{Pilbratt} G.~L. {et~al.}, 2010, \aap, 518, L1

\bibitem[{{Pineda}, {Langer} \& {Goldsmith}(2014){Pineda}, {Langer}, \&
  {Goldsmith}}]{Pineda2014}
{Pineda} J.~L., {Langer} W.~D., {Goldsmith} P.~F., 2014, \aap, 570, A121

\bibitem[{{Poelman} \& {Spaans}(2005)}]{Poelman2005}
{Poelman} D.~R., {Spaans} M., 2005, \aap, 440, 559

\bibitem[{{Poelman} \& {Spaans}(2006)}]{Poelman2006}
---, 2006, \aap, 453, 615

\bibitem[{{Popping}, {Behroozi} \& {Peeples}(2015){Popping}, {Behroozi}, \&
  {Peeples}}]{Popping2015}
{Popping} G., {Behroozi} P.~S., {Peeples} M.~S., 2015, \mnras, 449, 477

\bibitem[{{Popping} {et~al}\mbox{.}(2012){Popping}, {Caputi}, {Somerville}, \&
  {Trager}}]{popping2012}
{Popping} G., {Caputi} K.~I., {Somerville} R.~S., {Trager} S.~C., 2012, \mnras,
  425, 2386

\bibitem[{{Popping} {et~al}\mbox{.}(2015){Popping}, {Caputi}, {Trager},
  {Somerville}, {Dekel}, {Kassin}, {Kocevski}, {Koekemoer}, {Faber},
  {Ferguson}, {Galametz}, {Grogin}, {Guo}, {Lu}, {Wel}, \&
  {Weiner}}]{Popping2015Candels}
{Popping} G. {et~al.}, 2015, \mnras, 454, 2258

\bibitem[{{Popping} {et~al}\mbox{.}(2014){Popping}, {P{\'e}rez-Beaupuits},
  {Spaans}, {Trager}, \& {Somerville}}]{Popping2014radtran}
{Popping} G., {P{\'e}rez-Beaupuits} J.~P., {Spaans} M., {Trager} S.~C.,
  {Somerville} R.~S., 2014, \mnras, 444, 1301

\bibitem[{{Popping}, {Somerville} \& {Trager}(2014){Popping}, {Somerville}, \&
  {Trager}}]{Popping2014sam}
{Popping} G., {Somerville} R.~S., {Trager} S.~C., 2014, \mnras, 442, 2398

\bibitem[{{Porter} {et~al}\mbox{.}(2014){Porter}, {Somerville}, {Primack}, \&
  {Johansson}}]{Porter2014}
{Porter} L.~A., {Somerville} R.~S., {Primack} J.~R., {Johansson} P.~H., 2014,
  \mnras, 444, 942

\bibitem[{{Price}, {Federrath} \& {Brunt}(2011){Price}, {Federrath}, \&
  {Brunt}}]{Price2011}
{Price} D.~J., {Federrath} C., {Brunt} C.~M., 2011, \apjl, 727, L21

\bibitem[{{Prochaska} \& {Wolfe}(2009)}]{Prochaska2009}
{Prochaska} J.~X., {Wolfe} A.~M., 2009, \apj, 696, 1543

\bibitem[{{Robitaille} \& {Whitney}(2010)}]{Robitaille2010}
{Robitaille} T.~P., {Whitney} B.~A., 2010, \apjl, 710, L11

\bibitem[{{Rosenberg} {et~al}\mbox{.}(2014{\natexlab{a}}){Rosenberg},
  {Kazandjian}, {van der Werf}, {Israel}, {Meijerink}, {Wei{\ss}},
  {Requena-Torres}, \& {G{\"u}sten}}]{Rosenberg2014b}
{Rosenberg} M.~J.~F., {Kazandjian} M.~V., {van der Werf} P.~P., {Israel} F.~P.,
  {Meijerink} R., {Wei{\ss}} A., {Requena-Torres} M.~A., {G{\"u}sten} R.,
  2014{\natexlab{a}}, \aap, 564, A126

\bibitem[{{Rosenberg} {et~al}\mbox{.}(2014{\natexlab{b}}){Rosenberg},
  {Meijerink}, {Israel}, {van der Werf}, {Xilouris}, \&
  {Wei{\ss}}}]{Rosenberg2014a}
{Rosenberg} M.~J.~F., {Meijerink} R., {Israel} F.~P., {van der Werf} P.~P.,
  {Xilouris} E.~M., {Wei{\ss}} A., 2014{\natexlab{b}}, \aap, 568, A90

\bibitem[{{Rosenberg} {et~al}\mbox{.}(2015){Rosenberg}, {van der Werf},
  {Aalto}, {Armus}, {Charmandaris}, {D{\'{\i}}az-Santos}, {Evans}, {Fischer},
  {Gao}, {Gonz{\'a}lez-Alfonso}, {Greve}, {Harris}, {Henkel}, {Israel},
  {Isaak}, {Kramer}, {Meijerink}, {Naylor}, {Sanders}, {Smith}, {Spaans},
  {Spinoglio}, {Stacey}, {Veenendaal}, {Veilleux}, {Walter}, {Wei{\ss}},
  {Wiedner}, {van der Wiel}, \& {Xilouris}}]{Rosenberg2015}
{Rosenberg} M.~J.~F. {et~al.}, 2015, \apj, 801, 72

\bibitem[{{Sch{\"o}ier} {et~al}\mbox{.}(2005){Sch{\"o}ier}, {van der Tak}, {van
  Dishoeck}, \& {Black}}]{Schoier2005}
{Sch{\"o}ier} F.~L., {van der Tak} F.~F.~S., {van Dishoeck} E.~F., {Black}
  J.~H., 2005, \aap, 432, 369

\bibitem[{{Sharon} {et~al}\mbox{.}(2013){Sharon}, {Baker}, {Harris}, \&
  {Thomson}}]{Sharon2014}
{Sharon} C.~E., {Baker} A.~J., {Harris} A.~I., {Thomson} A.~P., 2013, \apj,
  765, 6

\bibitem[{{Somerville} \& {Dav{\'e}}(2015)}]{Somerville2014}
{Somerville} R.~S., {Dav{\'e}} R., 2015, \araa, 53, 51

\bibitem[{{Somerville} {et~al}\mbox{.}(2012){Somerville}, {Gilmore}, {Primack},
  \& {Dom{\'{\i}}nguez}}]{Somerville2012}
{Somerville} R.~S., {Gilmore} R.~C., {Primack} J.~R., {Dom{\'{\i}}nguez} A.,
  2012, \mnras, 423, 1992

\bibitem[{{Somerville} {et~al}\mbox{.}(2008){Somerville}, {Hopkins}, {Cox},
  {Robertson}, \& {Hernquist}}]{Somerville2008}
{Somerville} R.~S., {Hopkins} P.~F., {Cox} T.~J., {Robertson} B.~E.,
  {Hernquist} L., 2008, \mnras, 391, 481

\bibitem[{{Somerville}, {Popping} \& {Trager}(2015){Somerville}, {Popping}, \&
  {Trager}}]{Somerville2015}
{Somerville} R.~S., {Popping} G., {Trager} S.~C., 2015, \mnras, 453, 4337

\bibitem[{{Somerville} \& {Primack}(1999)}]{Somerville1999}
{Somerville} R.~S., {Primack} J.~R., 1999, \mnras, 310, 1087

\bibitem[{{Somerville}, {Primack} \& {Faber}(2001){Somerville}, {Primack}, \&
  {Faber}}]{Somerville2001}
{Somerville} R.~S., {Primack} J.~R., {Faber} S.~M., 2001, \mnras, 320, 504

\bibitem[{{Spaans} \& {Meijerink}(2008)}]{Spaans2008}
{Spaans} M., {Meijerink} R., 2008, \apjl, 678, L5

\bibitem[{{Swinbank} {et~al}\mbox{.}(2012){Swinbank}, {Karim}, {Smail},
  {Hodge}, {Walter}, {Bertoldi}, {Biggs}, {de Breuck}, {Chapman}, {Coppin},
  {Cox}, {Danielson}, {Dannerbauer}, {Ivison}, {Greve}, {Knudsen}, {Menten},
  {Simpson}, {Schinnerer}, {Wardlow}, {Wei{\ss}}, \& {van der
  Werf}}]{Swinbank2012}
{Swinbank} A.~M. {et~al.}, 2012, \mnras, 427, 1066

\bibitem[{{Tacconi} {et~al}\mbox{.}(2010){Tacconi}, {Genzel}, {Neri}, {Cox},
  {Cooper}, {Shapiro}, {Bolatto}, {Bouch{\'e}}, {Bournaud}, {Burkert},
  {Combes}, {Comerford}, {Davis}, {Schreiber}, {Garcia-Burillo},
  {Gracia-Carpio}, {Lutz}, {Naab}, {Omont}, {Shapley}, {Sternberg}, \&
  {Weiner}}]{Tacconi2010}
{Tacconi} L.~J. {et~al.}, 2010, \nat, 463, 781

\bibitem[{{Tacconi} {et~al}\mbox{.}(2013){Tacconi}, {Neri}, {Genzel}, {Combes},
  {Bolatto}, {Cooper}, {Wuyts}, {Bournaud}, {Burkert}, {Comerford}, {Cox},
  {Davis}, {F{\"o}rster Schreiber}, {Garc{\'{\i}}a-Burillo}, {Gracia-Carpio},
  {Lutz}, {Naab}, {Newman}, {Omont}, {Saintonge}, {Shapiro Griffin}, {Shapley},
  {Sternberg}, \& {Weiner}}]{Tacconi2013}
---, 2013, \apj, 768, 74

\bibitem[{{Tunnard} \& {Greve}(2016)}]{Tunnard2016}
{Tunnard} R., {Greve} T.~R., 2016, \apj, 819, 161

\bibitem[{{Vallini} {et~al}\mbox{.}(2016){Vallini}, {Gruppioni}, {Pozzi},
  {Vignali}, \& {Zamorani}}]{Vallini2016}
{Vallini} L., {Gruppioni} C., {Pozzi} F., {Vignali} C., {Zamorani} G., 2016,
  \mnras, 456, L40

\bibitem[{{van der Werf} {et~al}\mbox{.}(2010){van der Werf}, {Isaak},
  {Meijerink}, {Spaans}, {Rykala}, {Fulton}, {Loenen}, {Walter}, {Wei{\ss}},
  {Armus}, {Fischer}, {Israel}, {Harris}, {Veilleux}, {Henkel}, {Savini},
  {Lord}, {Smith}, {Gonz{\'a}lez-Alfonso}, {Naylor}, {Aalto}, {Charmandaris},
  {Dasyra}, {Evans}, {Gao}, {Greve}, {G{\"u}sten}, {Kramer},
  {Mart{\'{\i}}n-Pintado}, {Mazzarella}, {Papadopoulos}, {Sanders},
  {Spinoglio}, {Stacey}, {Vlahakis}, {Wiedner}, \& {Xilouris}}]{vanderWerf2010}
{van der Werf} P.~P. {et~al.}, 2010, \aap, 518, L42

\bibitem[{{Walter} {et~al}\mbox{.}(2014){Walter}, {Decarli}, {Sargent},
  {Carilli}, {Dickinson}, {Riechers}, {Ellis}, {Stark}, {Weiner}, {Aravena},
  {Bell}, {Bertoldi}, {Cox}, {Da Cunha}, {Daddi}, {Downes}, {Lentati},
  {Maiolino}, {Menten}, {Neri}, {Rix}, \& {Weiss}}]{Walter2014}
{Walter} F. {et~al.}, 2014, \apj, 782, 79

\bibitem[{{Wei{\ss}}, {Walter} \& {Scoville}(2005){Wei{\ss}}, {Walter}, \&
  {Scoville}}]{Weiss2005}
{Wei{\ss}} A., {Walter} F., {Scoville} N.~Z., 2005, \aap, 438, 533

\bibitem[{{White}, {Somerville} \& {Ferguson}(2015){White}, {Somerville}, \&
  {Ferguson}}]{White2014}
{White} C.~E., {Somerville} R.~S., {Ferguson} H.~C., 2015, \apj, 799, 201

\bibitem[{{Wolfire}, {Hollenbach} \& {McKee}(2010){Wolfire}, {Hollenbach}, \&
  {McKee}}]{Wolfire2010}
{Wolfire} M.~G., {Hollenbach} D., {McKee} C.~F., 2010, \apj, 716, 1191

\end{thebibliography}

\end{document}